\pdfoutput=1

\documentclass{article}

\usepackage{arxiv}
\usepackage{booktabs}
\usepackage[table,xcdraw]{xcolor}
\usepackage[utf8]{inputenc} %
\usepackage[T1]{fontenc}    %
\usepackage{hyperref}       %
\usepackage{url}            %
\usepackage{booktabs}       %
\usepackage{amsfonts,mathtools,amssymb,amsthm, amsmath}       %
\usepackage{array}
\newcolumntype{P}[1]{>{\centering\arraybackslash}p{#1}}
\usepackage{enumitem}
\usepackage{nicefrac}       %
\usepackage{microtype}      %
\usepackage{lipsum}
\usepackage{todonotes}
\usepackage[ruled]{algorithm2e} %
\usepackage[nameinlink,capitalize]{cleveref}
\newcommand{\A}{\mathcal{A}}
\newcommand{\V}{\mathcal{V}}

\newcommand{\R}{\mathbb{R}}

\newcommand{\bc}{\boldsymbol{c}}
\newcommand{\bt}{\boldsymbol{t}}
\newcommand{\bd}{\boldsymbol{d}}

\newcommand{\bh}{\boldsymbol{h}}
\newcommand{\by}{\boldsymbol{y}}
\newcommand{\bDelta}{\boldsymbol{\Delta}}

\newcommand{\bff}{\boldsymbol{f}}
\newcommand{\bFF}{\boldsymbol{F}}

\newcommand{\mR}{\mathcal{R}}

\DeclareMathOperator*{\argmin}{arg\min}

\usepackage[export]{adjustbox}
\usepackage{subcaption}
\usepackage{multirow}
\usepackage{rotating}
\usepackage{tikz}
\usetikzlibrary{positioning}
\usepackage{tikz-qtree}
\usetikzlibrary{arrows.meta}
\usepackage{subcaption}

\title{Quasi-Dynamic Traffic Assignment using High Performance Computing}
\author{
Cy Chan\thanks{Research Scientist at Lawrence Berkeley National Laboratory} \\
Computational Research Division\\
Lawrence Berkeley National Laboratory\\
Berkeley, CA 94720 \\
\texttt{cychan@lbl.gov} \\
\And
Anu Kuncheria\thanks{Graduate Student Researcher at University of California, Berkeley}\\
Institute of Transportation Studies\\
University of California\\
Berkeley, CA 94720 \\
\texttt{anu\_kuncheria@berkeley.edu} \\
\And
Bingyu Zhao\thanks{Researcher at University of California, Berkeley} \\
Institute of Transportation Studies\\
University of California\\
Berkeley, CA 94720 \\
\texttt{bz247@berkeley.edu } \\
\And
Theophile Cabannes\thanks{Graduate Student Researcher at University of California, Berkeley} \\
Electrical Engineering and Computer Science\\
University of California\\
Berkeley, CA 94720 \\
\texttt{theophile@berkeley.edu} \\ 
\And
Alexander Keimer\thanks{Researcher at University of California, Berkeley}\\
Institute of Transportation Studies\\
University of California\\
Berkeley, CA 94720 \\
\texttt{keimer@berkeley.edu} \\
\And
Bin Wang\thanks{Researcher at Lawrence Berkeley National Laboratory} \\
Energy Technology Area\\
Lawrence Berkeley National Laboratory\\
Berkeley, CA 94720 \\
\texttt{wangbin@lbl.gov} \\
\And
Alexandre Bayen\thanks{Professor at University of California, Berkeley}\\
Lawrence Berkeley National Laboratory\\
Electrical Engineering and Computer Science  \\
Institute of Transportation Studies \\
University of California\\
Berkeley, CA 94720 \\
\texttt{bayen@berkeley.edu} \\
\And
Jane Macfarlane\thanks{Executive Director Smart Cities at University of California, Berkeley} \\
Institute of Transportation Studies\\
Lawrence Berkeley National Lab\\
University of California\\
Berkeley, CA 94720 \\
\texttt{jfmacfarlane@lbl.gov}
}

\begin{document}
\maketitle

\begin{abstract}
Traffic assignment methods are some of the key approaches used to model the routing and flow patterns that arise in transportation networks. Since static traffic assignment does not have a notion of time, it is not designed to represent temporal dynamics that arise as vehicles flow through the network and demand varies throughout the day. Dynamic traffic assignment methods attempt to resolve these issues, but require significant computational resources if modeling urban-scale regions (on the order of millions of links and vehicle trips) and often take days of compute time to complete the optimization. 
The focus of this work is two-fold: 
1) to introduce a new traffic assignment approach - a quasi-dynamic traffic assignment (QDTA) model and 
2) to describe how we parallelized the QDTA algorithms to leverage High-Performance Computing (HPC) capabilities and scale to large metropolitan areas while dramatically reducing compute time.
We examine and compare different scenarios, including a baseline static traffic assignment (STA) and a quasi-dynamic scenario inspired by the user-equilibrium (UET).
Results are presented for the San Francisco Bay Area which accounts for about 19M trips/day and an urban road network of about 1M links and is validated against multiple data sources. 
We utilize an iterative gradient descent method, where the step size is selected using a Quasi-Newton method with parallelized cost function evaluations and compare it to using pre-defined step sizes (MSA).  Using the parallelized line search provides a 16 percent reduction in total execution time due to a reduction in the number of gradient descent iterations required for convergence.
The full day QDTA comprising 96 optimization steps over 15 minute intervals runs in about 4 minutes on 1,024 compute cores of the NERSC Cori computer, which represents a speedup of over 36x versus serial execution. 
To our knowledge, this compute time is significantly lower than other traffic assignment solutions for a problem of this scale.

\end{abstract}

\section{Introduction}
Transportation planners often use static traffic assignment solutions to estimate traffic states over the course of one day in their cities \cite{tsanakas_estimating_2020}.  Static traffic assignment does not deal with the dynamic behavior (realistically) that results from network dynamics - it simply assigns an origin/destination (O/D) routing solution that minimizes auto travel time for all mobile entities so that no drivers can unilaterally reduce his/her auto travel costs by shifting to another route. This is known as static user equilibrium (or Nash equilibrium, Wardrop's first principle) \cite{bell97transportation_networks}. To accommodate the temporal changes in the demand profile over an entire day, the problem is typically partitioned into time slots of interest and static traffic assignment solutions are used to estimate average speeds and average flows for each time slot for the network~\cite{scag_demand_model_2012}. Example time slots are early morning, morning rush hour, mid-day, evening rush hour, late evening - accounting for two to four hours of time in each time segment.  For very large-scale networks, these models often take many days to run depending upon the hardware and software solutions available to city planners.  Consequently, planners will often make adjustments to the model to make the computation tractable within their compute capabilities and time to solution requirements. For example, they may remove lower functional class roads from the network and aggregate travel demand to higher functional class roads. The results are then compromised by the loss of some roads and potential overestimation of flows on the remaining roads~\cite{flugel_traffic_2015}. 
Several limitations of STA has also been identified and discussed in many studies \cite{flugel_traffic_2015,tsanakas_estimating_2020,bliemer_requirements_nodate}. The underlying assumptions of STA, related to the stationary demand, long aggregation intervals, and static network loading, can lead to unrealistic results when variations in traffic conditions are high \cite{tsanakas_estimating_2020}. 

The focus of this work is to use static traffic assignment algorithms implemented on high-performance computing (HPC) to generate results that more effectively estimates a dynamic traffic assignment. We introduce a Quasi Dynamic Traffic Assignment (QDTA), which assigns traffic demand over a sequence of time intervals, where the traffic demand varies and the traffic flow propagates across intervals.  At each time step, a modified STA optimization problem is solved, which can be addressed by projected gradient methods where the gradient can be interpreted as a shortest path assignment.  We capture the dynamic impact of trips that span multiple time segments by introducing two mechanisms: path truncation and residual demand.  This approach addresses the complexity of a dynamic traffic assignment and removes the concern that static traffic assignment cannot provide value to practical traffic flow due to its single time assignment characteristic. In general, while the continuous-time DTA models satisfy the requirements of traffic flow theory, its uniqueness of network equilibrium flows is not necessarily guaranteed \cite{iryo_multiple_2011}. On the other hand, the quasi-dynamic model is based on static traffic assignment, somewhat sharing the formulation and properties of network equilibrium.  It is also referred to as discrete-time DTA or semi-dynamic traffic assignment in some literature \cite{bliemer_genetics_2017}. QDTA provides a better understanding of the traffic dynamics over STA as it can propagate traffic flows between the time periods using the residual demand. 

We use high-performance computing to accelerate the quasi-dynamic traffic assignment algorithm to generate high-fidelity traffic assignments with significantly improved computational efficiency.  Optimization on distributed-memory platforms is achieved by utilizing efficient routing algorithms and parallelizing multiple components of the computational workload across threads, including: 1) trip demand routing, 2) cost function evaluation, 3) network flow and weight updates, and 4) the residual demand calculations.  The improvements gained using HPC allow us to shorten the time segment duration to 15 minute intervals and address more complex road networks and increased travel demand profiles.  Our computational approach scales to one thousand compute cores, allowing us to run assignments with 19 million vehicle trips on a network with 1 million links in under 4 minutes.  This capability enables urban-scale traffic assignments in which a variety of scenarios can be addressed with less concern for computational complexity. Examples include: comparisons among counties or cities, integrating parallel discrete event simulation scenarios that implement and explore infrastructure implications of QDTA, and other objective functions for the optimization.

The remainder of this paper will be organized as follows.  First, we will discuss the general form of the proposed quasi dynamic traffic assignment in its pure mathematical form and introduce a proper instantiation of these class of problems, based on the well-known static traffic assignment. Second, we provide the QDTA models computational flow and algorithmic representations that estimate the dynamics using a sequence of time intervals. Third, we present the parallelization of the QDTA model and demonstrate the significant computational gains associated with this approach. Finally, we discuss the scenario comparison for the STA and QDTA models for the entire urban region of the San Francisco Bay Area.

\section{Quasi-dynamic traffic assignment and high-performance computing in the literature}

In the literature, the dynamics of QDTA comes in two components: demand/routing and dynamic/quasi-dynamic network loading.  QDTA differs from STA by the use of much smaller assignment intervals. The implication of having this temporal dimension is the presence of residual demand. The residual demand is handled differently in various previous work. \cite{bliemer2014quasi, tajtehranifard2018path} uses a point-queue model to include "capacity constraints", preventing flow from exiting the link if larger than the capacity. This gives more realistic travel times; however, the authors did not demonstrate its application in multi time step situations.

\begin{sidewaystable}
    \centering
    \caption{QDTA in the literature}
    \begin{tabular}{p{0.5cm}|p{3.5cm}|p{3.5cm}|p{3.0cm}|p{3.0cm}|p{3.0cm}|p{3.5cm}}
        \toprule
         & Routing & Network loading & Residual demand & Advantages & Limitations & Case studies  \\
         \midrule
         
        \cite{bliemer2014quasi} & Route-based SUE formulated as VI solved using MSA & first-order node model (i.e., reduction factor based on link demand and supply) formulated as a fixed-point problem solved iteratively & ... & Models node capacity at diverges and merges; correct representation of congestion upstream of bottleneck & Does not have temporal dynamics (\cite{bliemer2014quasi} distinguishes quasi-dynamic assignment from semi-static assignment) & Gold coast, Australia: 9,565 links, 2,987 nodes, 0.3 hour on a personal computer. Sydney, Australia: 75,379 links, 30,573 nodes, 1.1 hour on a personal computer \\
        
        \cite{tajtehranifard2018path} & System Optimal based on a set of reasonable and independent paths (path-based), assigned using MSA & same as above & ... & ... & ... & Sinox Falls \\
        
        \cite{fusco2013quasi} & SUE, from a shortest path set, no rerouting & Link cost function (e.g., BPR function) & ... & ... & No rerouting, limited path set & Rome, Italy: 15,000 links, 6,000 nodes, simulates several hours of traffic in a few minutes on a personal computer \\
        
        \cite{nakayama2014quasi} & & & & & & \\
        
        \cite{zeng2021data} & Dijkstra shortest path & Link cost constructed from average of GPS speed measurements, weighted by traffic volume reconstructed from the Licence Plate Recognition (LPR) data & No residual demand, likely the OD data is pre-partitioned & Data-driven link cost and demand & Not applicable where the high resolution data (e.g., GPS or toll demand) is limited (e.g., outside of the highways) & Expressway network in Hunan, China. 530 edges and 490 vertexes. Total length is 6725.5km. Computational performance unclear. \\
        
        \bottomrule
    \end{tabular}
    \label{tab:qdta_lit_review}
\end{sidewaystable}

\subsection{Dynamic traffic assignment and routing}
In this section, we give a short survey of existing modeling and routing approaches. Due to the vast number of publications in this area of research, the presented list of publications cannot be complete. However, we aim to describe the different approaches specifically and at a technical level for each class. We begin this overview with the famous books in this fields: \cite{chen2012dynamic,nagurney2012projected,ran2012dynamic}.

\paragraph{Continuous time modeling}
Generally, in a time-continuous setting, the link dynamics can be described via ODEs (ordinary differential equations) or PDEs (partial differential equations).
The articles \cite{merchant1978model,merchant,ran1989general,ran1993new,friesz1989dynamic} consider dynamic traffic assignment as optimal control problem over a given time horizon. The link dynamics are represented by a (system) of ODEs with in- and outflow functions. An objective function is specified and optimality conditions are deduced.
However, there is no natural delay and the optimization is considered on the full time horizon, so that one has to know information about inflow over the full time horizon to determine the routing, i.e.\ traffic assignment, not to mention that ODEs are generally quite far away from modeling traffic physically accurately (congestion patterns, the named delay, and more).

In \cite{ran1996link,boyce1995solving,friesz1993variational}, more general classes of ODE models (sometimes also with delay) are considered and time-dependent variational inequalities are determined to describe the routing at the junctions. Depending on whether the variational inequality is considered over the full time horizon or at every time (with the ordinary scalar product in \(\R^{n}\) and the variational inequality to hold at every time), the computation of a solution again requires full information of the input datum over the full time horizon considered.
In \cite{ran2012dynamic} a more detailed mathematical analysis concerning existence of solutions is provided.
\cite{bayen2019time} considers the well-posedness for an ODE model with delay and routing operators at every intersection which can dependent on the entire network state up to real-time.
In \cite{peeta2003stability} the routing is realized by a specific routing function taking into account the status of the network. It also proves some stability estimates with regard to the routing considered.
Another modeling approach which is also considered in \cite{ma2014continuous,ban2012modeling} consists of using the Vickrey or point-queue model (or a modified version) \cite{han,han_2,han_3}, again aiming for a routing (and departure choice time) based on variational inequalities.
These modeling approaches can be generalized to dynamics which are prescribed by \textit{partial differential equations} (PDEs). We refer to \cite{bressan2015optima,bressan2015conservation,garavello2016,garavello2006traffic,holden,bretti} for an overview. Usually, the underlying link dynamics are modeled by the LWR (Lighthill, Whitham, Richards) PDE \cite{lwr_1,lwr_2}, a hyperbolic conservation law, allowing spillback and congestion. Another approach uses non-local PDE models to simulate traffic flow at junctions \cite{laurent,keimer2,pflug3} and there are also higher order models available, see for instance \cite{aw2000resurrection,zhang2002non}. However, as mentioned before, a reasonable routing at the intersection needs to be prescribed and the underlying models need to be solved for a given space-time discretization on the entire network, resulting in a quite expensive computational situation.

As stated before, the (time dependent) routing itself -- which might depend on the entire network state at a given time (or also the future) -- will make the problem even more computationally challenging, not to mention the problem of calibrating the entire system reasonably.
These are reasons why we consider in this work a class of simpler models which still have a notion of time and delay, but not to the detail level which the previously mentioned models provide. Roughly, the model parameters required for a static traffic assignment are enough for the proposed QDTA.

\paragraph{Discrete time modeling}
There is a significant number of articles which deal with discrete time dynamical models. For the sake of an exhaustive overview, see \cite{ghali1995model,peeta1995system,smith1993new,waller2006chance,chen1998model,long}. 
The articles \cite{janson1991dynamic,janson1991convergent,ho1980successive} consider a discretized traffic flow model based on ODEs and time optimal control problem to obtain the routing. Most of the existing literature considers the assignment as an optimization problem across the entire period of interest, with the objective being either minimizing the total travel time (i.e., the System-Optimum, SO, condition, \cite{ho1980successive}), or some equivalence of the User Equilibrium (UE) conditions (\cite{nakayama2014quasi}). Depending on the decision variable used in the optimization formulation, the models can be categorized into link-based, path-based, or occasionally intersection-movement-based (\cite{long}). Link-based models can avoid enumerating and storing large path sets and are particularly advantageous for large networks.

In the discretized formulation, the choice of the time step interval is important. Generally, it should be much longer than the link travel time (\cite{janson1991convergent}), but not too long that makes each time slice close to static assignment. In addition, the size of the time step sometimes also depend on the availability of the temporal demand data, which may be available only at coarse grain (\cite{nakayama2014quasi}). The time-dependent demand is usually considered fixed and known, though \cite{waller2006chance} proposed algorithms for the cases with uncertainty in the demands (random with certain probability distributions). An important piece of consideration in time discretized DTA formulation is the treatment of residual demand, namely the trips that did not reach the destination in the previous time step. In the existing literature, the residual demand has been treated as model constraints (solved endogenously, e.g., in \cite{ho1980successive}), a fixed ratio over the inflow (\cite{nakayama2014quasi}), or obtained from a simulator (e.g., the DYNASMART simulator in \cite{peeta1995system}).

\subsection{High-performance computing for transportation modeling}

Parallel computation can be divided broadly into two categories: \textit{shared memory} (e.g. a workstation or a single server node) and \textit{distributed memory} (e.g. clusters, cloud platforms, and HPC systems)~\cite{correa2013models}.  The key distinction is that all cores in a shared memory system have direct access to data in the same address space.  That is, data that is written by each compute core is immediately visible to all other cores in the same memory.  In contrast, distributed memory systems have separate address spaces that require explicit message passing to share data between cores connected to distinct memory spaces.  Programming distributed memory systems is more difficult due to the separate memory, as it requires mechanisms to handle data and load distribution, synchronization, and data movement between processes.  Furthermore, there are increased performance overheads due to the time needed to communicate data and synchronize computation across processes.  However, the benefits of distributed memory parallelism are two-fold: access to more compute cores and a greater total memory capacity to enable solving larger problems more efficiently.  In this contribution, we consider the more difficult problem of distributed memory parallelization of the QDTA algorithm.

The use of high performance computing (HPC) in transportation modeling and simulation tools has not yet achieved widespread adoption.  Most simulation tools in this domain support either only sequential execution or (shared memory parallelism, which limits parallelism to the number of cores on a single compute node.  Examples that use this type of parallelism include the Aimsun~\cite{aimsun} and SUMO~\cite{sumo} simulators.  Some traffic simulation software projects, such as FastTrans~\cite{thulasidasan2009designing}, BEAM~\cite{beam} and POLARIS~\cite{auld2016polaris}, have also enabled the use of distributed memory systems.
Within the domain of traffic assignment, there have been some previous efforts that demonstrate the use of parallel computation on small to medium sized networks~\cite{petprakob2018implementation,himpe2019high}.  In \cite{petprakob2018implementation}, they analyze the Nagoya network consisting of 152K links and 38K nodes, using embarrassingly parallel path finding and processing only the active sub-networks to achieve up to 10x speedups using 25 processors.  In ~\cite{himpe2019high}, they utilize multi-threading and MPI to parallelize a link transmission model (LTM) based dynamic traffic assignment for a network with 11k nodes and 23K links, achieving up to 3x speedup on up to four cluster servers.  As we demonstrate in our contribution, the combination of an algorithm that lends itself to efficient parallelization along with the use of distributed computing presents a large opportunity to further increase the performance of traffic assignment algorithms for larger scale systems with \textbf{millions} of links and vehicles.

\section{A tractable dynamic assignment problem: QDTA formulation}\label{section:formulation}

The QDTA framework is to propose an efficient and parallelizable method to generate an optimized solution of traffic flows over the network according to the dynamic extensions of the widely used traffic assignment principles, such as the Wardrop's equilibrium, as shown in Sections \ref{section:residual_demand} and \ref{section:full_formulation}. The solution can be used to offer centralized and real-time routing guidance to enhance the traffic flow for a large network. Specifically, the Wardrop's equilibrium has two variations, namely the user equilibrium (UE) solution and the system optimal (SO) solution (see literature review). In this paper, the dynamic extension of the UE solution is adopted as an example in the mathematical formulation of the QDTA framework, with the detailed formulation given in Equation \ref{eq:path_flow_formulation} in Section \ref{section:full_formulation}.

The QDTA model proposed in this article divides the analysis period into small time steps and uses a sequence of STA steps to obtain an optimized route assignment for each time step. The main distinction between the proposed model and the conventional STA framework is the inclusion of \textit{route truncation} and \textit{residual demand}, to address the fact that some trip legs cannot be finished in a single analysis time step and need to be split across multiple steps.  As the analysis time step gets shorter to capture short-term traffic dynamics (i.e., 15 minutes), the fraction of trips spanning over multiple time steps become particularly prominent.  Section~\ref{sec:notation} introduces notation.  Section \ref{section:residual_demand} describes route truncation and residual demand in detail.  The integration of the residual demand into the well-understood framework of the STA is given in Section \ref{section:full_formulation}.  Lastly, some discussions are presented in Section \ref{section:qdta_math_discussions} regarding key questions such as the choice of time step length and the ability to resolve adaptive route selection (dynamic rerouting) behavior.

\subsection{Notation}
\label{sec:notation}

The road network is represented by a directed connected graph \(G=(\V, \A)\), where \(\V\) is the set of graph nodes/vertices and \(\A\) is the set of graph links/edges. A road intersection is represented by a node in the graph \(v \in \V\), and the stretch of road between two intersections is represented by a graph link \(a \in \A\). Each link has several static properties, such as the traffic flow capacity \(C_a\), and traversal time at the free-flow conditions (free-flow travel time) \(c_0\). Other static node and link properties, such as the coordinates of the nodes, the geometries of the links, are often stored (e.g., for visualization and post-processing analysis). But they are not described here since they are not integral in the formulation of the QDTA framework.

On the travel demand side, the time-dependent travel demand on this road network is represented by time-stamped origin-destination (OD) flows, \(d_{p, q}(i)\). \(p\) and \(q\) denote the origin and destination, with \(p \in \V\) and \(q \in \V\). Integer \(N\) is the total numbers of time steps considered.  The time values \(t_i \in \bt=\{t_0, t_1, ..., t_N\}\) refer to the bounds of the analysis time steps, and when we refer to time interval $i$, we are referring specifically to interval $[t_i, t_{i+1})$.  Thus, the duration of time interval $i$ is \(\Delta t_i = t_{i+1} - t_i, \ \forall \ 0 \le i < N \). For example, if considering a traffic assignment period of 24 hours with a uniform time interval of 15 minutes, \(N\) then equals to 96 and \(\Delta t_i\) is 15 minutes for all \(i\).
Let \(\mR_{p, q}\) represents the set of acyclic paths in the graph that connect \(p\) and \(q\).
The traffic flow on each path \(r \in \mR_{p, q}\) is denoted by \(h_r(i)\), or \(\bh(i)\) for all path flows at a particular time interval $i$.  Similarly, the traffic flow on each link \(a \in \A\) is denoted by \(f_a(i)\), or \(\bff(i)\) for all links.

In our discussion of the parallelized versions of the QDTA algorithms (Section~\ref{sec:parallelization}), we require notation to represent each compute thread's local view (i.e. partition or partial sum) of the above variables.  In the rest of this paper, we use the term \textit{partition} to mean the division of a parent vector into two or more child vectors in a way such that each element of the parent vector is assigned to exactly one child vector.  In this way, partitioning a parent vector into child vectors is analogous to splitting a parent set into disjoint subsets whose union is the parent.  We utilize the subscript $k \in \{1, 2, \ldots, M\}$, where $M$ is the total number of threads, to denote the $k$-th thread's partition of the original variable.  For example, parallel decomposition of the demand is accomplished by partitioning the total travel demand \(\bd(i)\) into \(\{ \bd_k(i) \  | \  \forall k \}\), and similar for \(\bd^o(i)\) and \(\bd^r(i)\).  Furthermore, \(\bh_k(i)\) represents the flows on the paths that correspond to \(\bd_k(i)\).  Another form of parallel decomposition is of the links in the network, where \(\A_k\) represents thread \(k\)'s partition of \(\A\).  \(\bff_k(i)\) is different in that it represents the flows on \textbf{all} links that result from the partitioned demand \(\bh_k(i)\), such that \(\bff(i) = \sum_k \bff_k(i)\).  These variables and their use will be further detailed in Section~\ref{sec:parallelization}.

\begin{table}
    \caption{Meanings of mathematical expressions}
    \centering
    \begin{tabular}{c p{10cm}}
        \toprule
        Symbol & \multicolumn{1}{c}{Meaning}  \\
        \midrule
        \multicolumn{2}{c}{\textbf{Parallel decomposition}} \\
        \(M, k \in \{1, 2, \ldots, M\}\) & Number of compute threads, thread index. \\
        \hline
        
        \multicolumn{2}{c}{\textbf{Time-step related variables}} \\
        \(N\), \(i \in \{0, 1, \ldots, N-1\} \) & Total number of time intervals, time interval index. \\
        \([t_i, t_{i+1})\) & \(i^{th}\) time interval bounds. \\
        \(\Delta t_i = t_{i+1} - t_i\) & Duration of time interval \(i\). \\
        \hline

        \multicolumn{2}{c}{\textbf{Network properties}} \\
        \(G = (\V, \A)\) & Road network graph. \\
        \(\V\), \(v\) & Set of road network vertices, one vertex in the road network. \\
        \(\A\), \(a\) & Set of road network edges, one edge in the road network. \\
        \(\A_k\) & Thread \(k\)'s partition of network edges. \\
        \(C_a\) & Flow capacity of edge \(a\). \\
        \(\bc_0\), \(c_{0,a}\) & Free-flow travel time/cost of all edges, or a specific edge \(a\). \\
        \(\bc(i)\), \(c_a(i)\) & Time/Cost of traversing all edges, or a specific edge \(a\), in time interval \(i\). \\
        \hline
        
        \multicolumn{2}{c}{\textbf{Travel demand}} \\
        \(p \in \V\) & Trip origin. \\
        \(q \in \V\) & Trip destination. \\
        \(s \in \V\) & Intermediate stop location. \\
        \(d_{p, q}(i)\) & Travel demand (trips) starting at node \(p\), ending at node \(q\), departing in time interval \(i\) . \\
        \(\bd^o(i)\), \(\bd^r(i)\), \(\bd(i)\) & All original, residual, and combined trips traveling in time interval \(i\). \\
        \(\bd_k^o(i)\), \(\bd_k^r(i)\), \(\bd_k(i)\) & Thread \(k\)'s partition of corresponding trips in time interval \(i\). \\
        \hline

        \multicolumn{2}{c}{\textbf{Paths and network flows}} \\
        \(\mR_{p,q}\), \(r\) & All acyclic paths in the road network that connect \(p\) and \(q\), a single path in the set. \\
        \(len(r)\), \(r_k\) & Total number of vertices along path \(r\), the \(k^{th}\) vertex along path \(r\). \\
        \(\bh(i)\), \(h_r(i)\) & Traffic flow assigned to all paths, or path \(r\), in time interval \(i\). \\
        \(\bff(i)\), \(f_a(i)\) & Traffic flow assigned to all links, or link \(a\), in time interval \(i\). \\
        \(\bh_k(i)\), \(\bff_k(i)\) & Traffic flow assigned to paths corresponding to \(\bd_k(i)\), and their resulting partial link flows. \\
        \bottomrule
    \end{tabular}
    \label{tab:math_symbols}
\end{table}

\subsection{Residual demand and route truncation}\label{section:residual_demand}

In QDTA, travel demands are supplied in stages (time step intervals) according to the analysis period that the departure times are in.
The assignment framework presented in this paper does not constrain the duration of the time steps. As a result, different time step lengths can be used, e.g., 5 minutes or 1 hour. However, a time step that is too long would approximate the STA solution and lack the desired temporal variations in the modeling results, while too short a time step increases computational cost and requires a higher resolution temporal distribution of the demand inputs.

A major complication of selecting a small time interval (relative to the duration of the average trip, e.g., 15 minutes) is that a large fraction of trips do not finish within one assignment period. This is problematic as established optimization algorithms to obtain traffic assignment solutions (e.g., the Frank-Wolfe's algorithm) assume all vehicles reach their destinations during the assignment~\cite{frank1956algorithm}, which is not realistic for assignment over short time intervals.  To address this issue, we introduce a modification to the STA approach by adding a \textit{route truncation} operation that is used repeatedly in the algorithm. In this section, the mathematical formulation of the route truncation operation will be introduced. Its integration into the QDTA framework will be given in Section \ref{section:full_formulation}.

The route truncation operation estimates the intermediate stop location \(s\) that a trip in \(d_{p,q}(i)\) can reach before time \( t_{i+1}\). For long trips, \(s\) may be different from the trip destination \(q\), and the remaining leg of the trip, i.e., from \(s\) to \(q\), will enter the next time step as the \textit{residual demand}. The determination of \(s\) relies on knowing the travel time (or other general cost) of each link, \(\bc(i)=\{c_a(i)\}, a\in\A\), which is a function of the link flow \(\bff(i)=\{f_a(i)\}, a\in\A\) assigned to the network during interval \(i\).

In general, the travel time costs are a function of the link flows: \(\bc(i) = \bc(\bff(i)), \forall i\).  This function is usually selected to satisfy both element-wise monotonicity and separability assumptions.  The element-wise monotonicity underlines the fact that as more flow is assigned to a road, the longer it takes on average to traverse it. The separability assumption states that travel time on a link only depends on the current flow assignment on the considered link.
A typical choice of the travel time function that satisfies both assumptions is the well-known Bureau of Public Roads (BPR) curves \cite{patriksson}, as shown in Equation (\ref{eq:link_travel_time}). In Equation (\ref{eq:link_travel_time}), \(c_a(i)\) is the travel time on link \(a\) in time interval \(i\); \(c_{0,a}\) and \(C_a\) are the free-flow travel time and capacity associated with the link; \(\alpha\) and \(\beta\) are calibration parameters selected such that $c_a(i)$ is monotonic with respect to $f_a(i)$.

\begin{equation}
c_a(i) = c_{0,a} \cdot \left(1+\alpha\Big(\tfrac{f_a(i)}{C_a}\Big)^{\beta}\right)
\label{eq:link_travel_time}
\end{equation}

Based on the link-level travel time \(\bc(i)\), the stop location \(s\) can be determined using Equation (\ref{eq:stop_node}). \(r\) is the path of the trip to its destination \(q\). If the time step duration is long enough, longer than the total time required to traverse \(r\), i.e., first row in Equation (\ref{eq:stop_node}), the intermediate stop location \(s\) is the trip destination \(q\). However, if this condition cannot be met, the stop location is a node along the path \(r\). Let \(len(r)\) denote the total number of nodes along path \(r\), and \(r_j\) to be the \(j^{th}\) node on the path, the second row of Equation (\ref{eq:stop_node}) leads to the furthest distance that can be covered during a time duration of \(\Delta t_i\).

\begin{equation}
    s = 
    \begin{cases}
        q & \text{if } \Delta t_i \geq \underset{a \in r}{\sum} c_a(i) \\
        r_{\underset{j \in len(r)}{\argmin} \; \underset{a \in r_j}{\sum} max\{\Delta t_i - c_a(i),0\}} & \text{otherwise} \\
    \end{cases}
    \label{eq:stop_node}
\end{equation}

Thus, for trip \(d_{p, q}(i)\) with route \(r\), only the first part of the trip up to vertex \(s\) will contribute towards the path/link flow in the time interval \(i\), while the remainder of the trip will be added to the next time step's in the form of carry-over demand \(d_{s, q, t_{i+1}}\).  Note that particularly long trips may carry over several times, and thus span several time segments.

\subsection{Full formulation} \label{section:full_formulation}

In this section, the temporal update steps to obtain the path/link traffic flow assignment will be presented. At each time step, the travel demand \(\bd(i)\) consists of two parts, the original trips \(\bd^o(i)\) that start in time interval \(i\), and the residual demand trips \(\bd^d(i)\) that started before \(t_i\) but have not yet reached their destinations. This applies to all time steps except the first time interval 0, where there is no residual demand. Let \(\bFF\) be the function that maps the travel demand to path flows:

\begin{equation}
  \bh(i) =
    \begin{cases}
        \bFF(\bd^o(0)) & i = 0 \\
        \bFF(\bd^o(i) + \bd^r(i)) & i \in {1, ..., N-1} \\
    \end{cases}
    \label{eq:path_flow_formulation}
\end{equation}

Then, Equation (\ref{eq:path_flow_formulation_ue}) is an instantiation of \(\bFF\) that finds an optimum static assignment of path flow \(\bh=\{h_{p,q,r}\}, p,q \in \V, r \in \mR_{p,q}\) that satisfies the UE condition (provided the cost function \(c_a(s)\) is strictly monotone) with proof in \cite{patriksson}.

\begin{subequations}
\begin{align}
    & \bFF(\bd) = \underset{\bh}{\argmin} \underset{a \in \A}{\sum} \int_{0}^{f_a} c_a(s) ds & \\
    \text{subject to}
    & \underset{r \in \mR_{p,q}}{\sum} h_{r} = d_{p,q} & \forall (p, q) \in \V \times \V \\
    & h_{r} \geq 0 & \forall r \in \mR_{p,q} \\
    & f_a = \sum_{p, q \in \V} \sum_{r \in \mR_{p,q}} \Delta_{a,r} h_{r} & a \in \A
\end{align}
\label{eq:path_flow_formulation_ue}
\end{subequations}

\(\bDelta = \left[ \Delta_{a,r} \right] \) is the incidence matrix, whose value is 1 if the link is on the path before the intermediate stop \(s\), and 0 otherwise:

\begin{equation}
    \Delta_{a,r} = 
    \begin{cases}
        1 & \text{if } a \in \{(r_0, r_1), ..., (r_{j-1}, r_j)\}, r_j = s \\
        0 & \text{otherwise} \\
    \end{cases}
    \label{eq:incident_matrix}
\end{equation}

Equation (\ref{eq:path_flow_formulation_ue}) is essentially an extension of the static UE formulation in \cite{patriksson}, with the static travel demand, link and path flow replaced by the time-dependent \(\bd(i)\), \(\bff(i)\) and \(\bh(i)\). Unlike the original static formulation, where the path-link incident matrix only depends on knowing whether a link is on a path, in the quasi-dynamic formulation, a link on a path may not be traversed in the current time step. As a result, the path flow in QDTA is truncated at the stop location, \(s\), first using Equation (\ref{eq:stop_node}), before mapped to the link flow \(\bff(i)\) using Equations (\ref{eq:path_flow_formulation_ue}d)-(\ref{eq:incident_matrix}).

\subsection{Discussions of the QDTA framework}\label{section:qdta_math_discussions}

Some further discussions are provided regarding the proposed QDTA framework. First, the selection of the time interval durations \(\{ \Delta t_i \}\) reflects a few trade-offs. If the time step length is too long, the solution will approximate the STA solution and fail to capture the changing temporal traffic dynamics. However, if the time step length is too short, it will not only increase the computational time significantly, but also lead to less accurate results as microscopic, sub-link dynamics start to show. For example, consider an extreme scenario, where \(\Delta t_i\) is shorter than the free flow time of the road link, \(c_{0,a}\), Equation (\ref{eq:stop_node}) indicates that the intermediate stop node \(s\) is always the origin node. In other words, the trips can never propagate through the link. This limitation can potentially be overcome by using alternative formulations of Equation (\ref{eq:stop_node}) that tracks the distance of the traffic flow on a link, and cumulatively adds up the distance across time steps if the traffic flow cannot be propagated to the next link.

A key feature of the proposed QDTA framework is the inclusion of dynamic rerouting in response to changes in the network at each time interval. For every time interval $i$, the routing is updated to reflect the current traffic conditions and potential changes in the network (e.g., road closures due to earthquake damages). The route assignment is thus obtained through an iterative process between the path flow, \(\bh(i)\), and the path truncation, and reflects the optimum (e.g., UE) traffic distribution within each time step. The dynamic rerouting is not considered in some previous DTA algorithms. But it is particularly relevant, especially in the era of navigation applications that dynamically update the navigation routes for users, to include such phenomena in traffic modeling and predictions.

\section{Algorithmic Representations}
\label{sec:algorithms}

The mathematical formulation of the QDTA framework in Section \ref{section:formulation} is solved using a number of scalable algorithms. The pseudo-code of these algorithms are presented in this section. At the highest level, the QDTA implementation (\cref{alg:qdta}) loops through each time step \(\{t_0,\ t_1,\ ...,\ t_{N-1}\}\) (\cref{alg:qdta}). Inside each time step, an STA-based flow solution is first obtained using the Frank-Wolfe's algorithm (\cref{alg:sta}, \cref{alg:aon}). Note that path truncation is performed in each iterative step of the Frank-Wolfe's algorithm (\cref{alg:aon}). This ensures that, for each path, only the approximate portion traversed {\it within the current time segment} is considered so as to prevent erroneous congestion effects of traffic that occurs outside the current time segment.  Based on the converged solution from \cref{alg:sta}, a last round of route truncation is performed to calculate the intermediate positions $s$ of the residual demand to be carried over to the next time step (\cref{alg:residual}).

\begin{algorithm}
\KwData{
    Network graph $G(\V, \A)$ \\
    \qquad \quad Time step length $\Delta t_i$ \\
    \qquad \quad Total time step counts $N$ \\
    \qquad \quad Original travel demand of all time steps $\bd^o = \{\bd^o(i)\}$, with $i \in \{0, ..., N-1\}$ \\
    }
\texttt{\\} 
Initialize $\bd^r(0) = $ empty nested associative array \tcp*{No residual demand in the first time step}
\texttt{\\} 
\For{$i \in \{0, \ldots, N-1 \}$ \tcp*{Sequential discrete time step}}{
 $\bd(i) = \bd^o(i) \mathrel{+} \bd^r(i)$ \tcp*{Get original and residual demand}
 $\bh^{STA}(i) = $ Traffic\_assignment$(G,\ \Delta t_i,\ \bd(i))$ \tcp*{Path flow assignment using STA}
 $\bh(i), \bd^r(i+1) = $ Residual\_demand$(G,\ \Delta t_i,\ \bh^{STA}(i))$ \tcp*{Truncate path flows and get residual demand based on time step length}
}
\KwResult{$\{\bh(i) \ | \ i \in \{0, \ldots, N-1\}\}$ \tcp*{Path flow for all time steps}}
 \caption{Quasi-dynamic traffic assignment}
 \label{alg:qdta}
\end{algorithm}

\begin{algorithm}
\KwData{ Network graph $G(\V, \A)$ \\
        \qquad \quad Time step length $\Delta t$ \\
        \qquad \quad Travel demand of current step $\bd \in \R_{\ge 0}^{\V\times\V}$ \\
        \qquad \quad Free-flow travel time of each link $\bc_0 = \{c_{0,a}\}$, with $a \in \A$ \\
    }
\texttt{\\} 
Take $\bh =$ All\_or\_nothing$(G, \Delta t, \bd, \bc_0)$ \tcp*{Set initial path flow in the free-flow condition}
\texttt{\\} 
\While{\textbf{True} \tcp*{Gradient descent step}}{
$\bc = \textrm{BPR}(\bDelta \bh)$ \tcp*{Calculate the edge travel time}
$\bh^{AON} =$ All\_or\_nothing$(G, \Delta t, \bd, \bc)$ \tcp*{All-or-nothing path flow with new edge weights}
$\alpha^\star = \argmin_\alpha$ Cost\_function$\left(\bDelta \left[\bh + \alpha\cdot(\bh^{AON}-\bh)\right]\right)$ \tcp*{Exact line search}
$\bh_{new} = \bh + \alpha^\star \cdot(\bh^{AON}-\bh)$ \tcp*{Update path flows}
\If{ \textrm{Converged} ($\bDelta \bh, \bDelta \bh_{new}$) \tcp*{Check convergence}} {
\textbf{break}
}
$\bh = \bh_{new}$ \;
}
\KwResult{$\bh$ \tcp*{Path flow using STA}}
 \caption{Traffic\_assignment: using Frank-Wolfe's algorithm}
 \label{alg:sta}
\end{algorithm}

\begin{algorithm}
\KwData{ Network graph $G(\V, \A)$ \\
        \qquad \quad Time step length $\Delta t$ \\
        \qquad \quad Travel demand of current step $\bd \in \R_{\ge 0}^{\V\times\V}$ \\
        \qquad \quad Edge travel time/cost $\bc = \{c_a\}, a \in \A$ }
\texttt{\\}
Initialize $\bh =$ empty associative array \tcp*{Initialize the path flow vector}
\texttt{\\} 
\For{$d_{p,q} \in \bd \ | \ d_{p,q} > 0$ \tcp*{Iterate over non-zero elements}}{
$r_{sp} = $ Get\_shortest\_path$(p, q, \bc, G)$  \tcp*{Get shortest path to destination}
$r_{tp} = $ Truncate\_path$(r_{sp}, \bc, G, \Delta t)$  \tcp*{Truncate path according to time step length}
$ h_{r_{tp}} \mathrel{+}= d_{p,q}$ \tcp*{Add trips to the path flow vector}
}
\KwResult{$\bh$ \tcp*{Path flow from all-or-nothing assignment}}
 \caption{All\_or\_nothing: iterative step of the Frank-Wolfe Algorithm}
 \label{alg:aon}
\end{algorithm}

\begin{algorithm}
\KwData{ Network graph $G(\V, \A)$ \\
        \qquad \quad Time step length $\Delta t_i$ \\
        \qquad \quad Intermediate path flow results from the STA $\bh^{STA}(i)$ \\
        \qquad \quad All paths used in the time step $\mR(i) = \{\mR_{p,q}\}$, with $(p,\ q) \in \R_{\ge 0}^{\V\times\V}$ 
    }
\texttt{\\} 
Initialize $\bd^r(i+1) =$ empty nested associative array \tcp*{Initialize the residual demand vector, will be added to the demand of the next step}
Initialize $\bh(i) =$ empty associative array \tcp*{Initialize the truncated path flow vector}
Take $\bc = BPR(\bDelta \bh^{STA}(i))$ \tcp*{Edge travel time based on path/link flow from STA}
\texttt{\\} 
\For{$r \in \mR(i)$ \tcp*{Parallelizable for loop}}{
$r_{tp} = $ Truncate\_path$(r, \bc, G, \Delta t_i)$ \tcp*{Truncate path to what can be traversed in the time step}
$h_{r_{tp}}(i) \ \mathrel{+}= h^{STA}_{r}(i)$ \tcp*{Populate the path flow vector}
\If{$r_{tp} \neq r$ \tcp*{Flow has not reached its destination within the time slice}}{
$s = $ Get\_last\_vertex$(r_{tp})$\;
$q = $ Get\_last\_vertex$(r)$\;
$d^r_{s,q}(i+1) \mathrel{+}= h^{STA}_{r}(i)$ \tcp*{Add to residual demand}
}
}
\KwResult{$\bh(i)$, $\bd^r(i+1)$ \tcp*{Path flow for the current time step using QDTA, and residual demand to be added to the next time step}}
 \caption{Residual\_demand: trips that cannot finish in one assignment interval}
 \label{alg:residual}
\end{algorithm}

The repeated usage of route truncation resolves the issue of traffic that is resident in the network for a longer time horizon than the given QDTA time segment duration by re-assigning the residual traffic to the next time interval.
With this capability, the demand entering the network at the next time interval along with the residual demand is then introduced to the following time segment's traffic assignment. 
This process is repeated until the final time of the QDTA is reached. 
The model can be interpreted as an time-expanded network approach where the expansion of the network represents the evolution with respect to time. 

In the algorithm, the following operations are used but not explicitly defined:
\begin{itemize}
    \item \verb Shortest_path  returns the shortest path between the origin $p$ and the destination $q$ given the road network graph $G(\V, \A)$ with cost $\bc = \{c_a\}$, with $a \in \A$.
    \item \verb Truncate_path  returns the sub-path of a path $r$ that can be traversed in time $\Delta t_i$ given the travel time on the edges.
    \item \verb Get_last_vertex  returns the last vertex of a path $r$.
    \item \verb Cost_function  is the function to be minimized, i.e. the Rosenthal potential function with $c_a(s)$ chosen to be a monotonic function as in \cref{eq:link_travel_time}:
        \begin{equation}
        \textrm{Cost\_function}(\bff) = \sum_{a \in \A} \int_0^{f_a} \bc_a(s) \ ds \label{eq:cost_function}
        \end{equation}
    \item \verb Converged  checks the convergence of the Frank-Wolfe algorithm. The algorithm is considered to be converged if the relative absolute change in the cost function is less than $T_0 = 0.0001$.  This value was chosen as a compromise between quality and speed of convergence.
        \begin{equation}
        \textrm{Converged}(\bff_1, \bff_2) = 
        \begin{cases}
            \textrm{True}, & \text{if } \left| \frac{ \textrm{Cost\_function}(\bff_1) - \textrm{Cost\_function}(\bff_2) }{\textrm{Cost\_function}(\bff_1)} \right| < T_0 \\
            \textrm{False}, & \text{otherwise}.
        \end{cases}
        \end{equation}
\end{itemize}

\section{High-performance computational solutions for QDTA}
\label{sec:hpc_qdta}

Our overall goal is to reduce the compute time for modeling urban mobility so that city planners can investigate a large number of scenarios in a reasonably small period of time, while still preserving the fidelity of the inherent traffic dynamics.
The scope of the models considered in our work are full urban networks that include all functional road classes and a full urban scale travel demand.
Our software framework takes the network graph (nodes and links) and traffic demand (origin and destination nodes and start times) as inputs.
The demand is converted into $\{\bd^o(i) \ | \ \forall i \}$ based on their origin and destination nodes $p, q \in \V$ and starting times $t$.

\begin{figure}
  \centering
   \includegraphics[width=\linewidth]{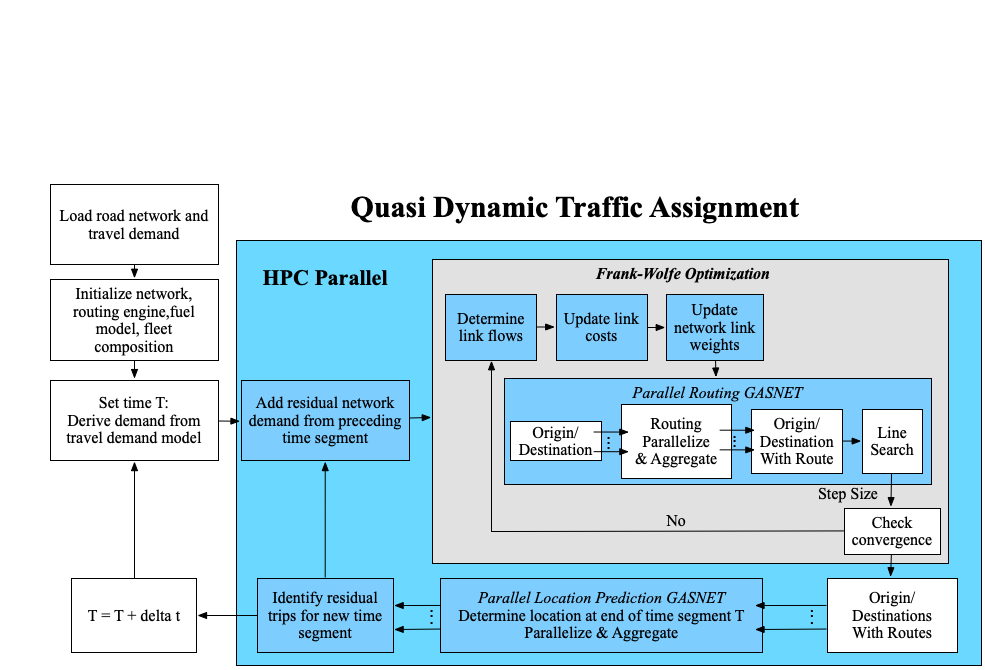}
  \caption{Computational Flow for QDTA : blue box indicates the parts which have been parallelized. This flow diagram corresponds to the pseudo-algorithm described in Algorithm 3.}
  \label{fig:QDTAFlowDiagram}
\end{figure}

\Cref{fig:QDTAFlowDiagram} shows the computational flow of our implementation of QDTA.  All algorithms in the blue boxes have been parallelized to run on HPC platforms.  The main (outer) loop in \Cref{fig:QDTAFlowDiagram} corresponds to one iteration of \cref{alg:qdta}.  The Frank-Wolfe optimization is fully parallelized, including the routing, line search, and network weight updates.  The method of parallelization of the core algorithms described in Section~\ref{sec:algorithms} will be described in detail in the remainder of this section.

\subsection{Routing and network flow parallelization}
\label{sec:parallelization}

\begin{algorithm}
\KwData{
    Network graph $G(\V, \A)$ \\
    \qquad \quad Time step length $\Delta t_i$ \\
    \qquad \quad Total time step counts $N$ \\
    \qquad \quad Original travel demand of all time steps $\bd^o = \{\bd^o(i)\}$, with $i \in \{0, ..., N-1\}$ \\
    }
\texttt{\\} 
Let $k$ be this thread's index \;
Initialize $\bd_k^r(0) = $ empty nested associative array \tcp*{No residual demand in the first time step}
\texttt{\\} 
\For{$i \in \{0, \ldots, N-1 \}$}{
 Let $\bd_k^o(i)$ be thread $k$'s partition of $\bd^o(i)$ \;
 $\bd_k(i) = \bd_k^o(i) \mathrel{+} \bd_k^r(i)$ \tcp*{Combine original and residual demand} 
 $(\bh_k^{STA}(i), \bff^{STA}(i)) = $ Parallel\_traffic\_assignment$(G,\ \Delta t_i,\ \bd_k(i))$ \;
 $\bh_k(i), \bd_k^r(i+1) = $ Parallel\_residual\_demand$(G,\ \Delta t_i,\ \bh_k^{STA}(i))$ \;
}
\KwResult{$\bh_k = {\bh_k(i)}$, with $i \in {0, ..., N-1}$ \tcp*{Distributed path flows for all time steps}}
 \caption{Parallel quasi-dynamic traffic assignment}
 \label{alg:parallel_qdta}
\end{algorithm}

Algorithm~\ref{alg:parallel_qdta} describes the distributed-memory parallel QDTA algorithm.  When parallelizing any algorithm, one of the design choices is how to partition both the data and computational work across available compute resources.  This choice may differ depending on the algorithmic step being parallelized (see Figure~\ref{fig:QDTAFlowDiagram}).  
Due to the high computational cost of computing shortest path routes over a large network, the most critical step to optimize for performance is the routing of all active vehicles in the current time segment (Algorithm~\ref{alg:aon}).  We can parallelize the routing with respect to each origin and destination pair $p, q \in \V$ as they do not influence each other within a single step of the Frank-Wolfe algorithm.  In order to achieve effective parallelization of the routing step, and to distribute storage and management of the resulting routes, each thread $k$ is assigned a partition of the demand $\bd_k^o(i)$ to compute their routes and flows $\bh_k(i)$, manage their corresponding residual demand $\bd_k^r(i+1)$, and finally store their results to disk at program completion.

\begin{algorithm}
\KwData{ Network graph $G(\V, \A)$ \\
        \qquad \quad Time step length $\Delta t$ \\
        \qquad \quad Partition of travel demand of current step $\bd_k$ \\
        \qquad \quad Free-flow travel time of each link $\bc_0 = \{c_{0,a}\}$, with $a \in \A$
    }
\texttt{\\} 
Let $(\bh_k, \bff) =$ Parallel\_all\_or\_nothing$(G, \Delta t, \bd_k, \bc_0)$ \tcp*{Set initial path flow}
\texttt{\\} 
\While{\textbf{True} \tcp*{Gradient descent step}}{
$\bc = BPR(\bff)$ \tcp*{Calculate the edge travel time}
$(\bh_k^{AON}, \bff^{AON}) =$ Parallel\_all\_or\_nothing$(G, \Delta t, \bd_k, \bc)$ \;
$\alpha^\star =$ Parallel\_line\_search$(G, \bff, \bff^{AON})$ \;
$\bff_{new} = \bff + \alpha^\star \cdot (\bff^{AON} - \bff)$ \tcp*{Update link flows}
\If{ \textrm{Converged} ($\bff, \bff_{new}$) \tcp*{Check convergence}} {
\textbf{break}
}
$\bh_k = \bh_k + \alpha^\star \cdot(\bh_k^{AON}-\bh_k)$ \tcp*{Update path flows}
$\bff = \bff_{new}$ \;
}
\KwResult{$\bh_k, \bff$ \tcp*{Path flows (local) and link flows (global)}}
 \caption{Parallel\_traffic\_assignment: using Frank-Wolfe's algorithm}
 \label{alg:parallel_sta}
\end{algorithm}

\begin{algorithm}
\KwData{ Network graph $G(\V, \A)$ \\
        \qquad \quad Time step length $\Delta t$ \\
        \qquad \quad Partition of travel demand of current step $\bd_k$ \\
        \qquad \quad Edge travel time/cost $\bc = \{c_a\}, a \in \A$ }
\texttt{\\}     
Pre-process (customize) network with updated link weights $\bc$ \;
Initialize $\bh_k =$ empty associative array \tcp*{Initialize the path flow vector}
\texttt{\\} 
\For{ $d_{p,q} \in \bd_k \ | \ d_{p,q} > 0$ \tcp*{Iterate over non-zero elements}} {
$r_{sp} = $ Get\_shortest\_path$(p, q, \bc, G)$  \tcp*{Get shortest path}
$r_{tp} = $ Truncate\_path$(r_{sp}, \bc, G, \Delta t)$  \tcp*{Truncate path according to time step length}
$ h_{r_{tp}} \mathrel{+}= d_{p,q}$ \tcp*{Add trips to the local path flow vector $\bh_k$}
}
$\bff_k = \bDelta \bh_k$ \tcp*{Compute local link flows from local path flows}
$\bff = $ Global all-reduce$(+, \bff_k)$ \tcp*{Compute global link flows from local link flows}
\KwResult{$\bh_k, \bff$ \tcp*{Path flows (local) and link flows (global)}}
 \caption{Parallel\_all\_or\_nothing: iterative step of the Frank-Wolfe Algorithm}
 \label{alg:parallel_aon}
\end{algorithm}

Algorithm~\ref{alg:parallel_sta} describes the parallelized Frank-Wolfe algorithm that assigns traffic to each time segment.  The first step in this algorithm is the parallel all-or-nothing routing step described in Algorithm~\ref{alg:parallel_aon}.  Note that each thread requires full knowledge of the network's current link weights to be able to route its partition of trip legs $d_k$.  An important optimization we made for the routing step involves utilizing a multi-phase routing algorithm, where the network is first pre-processed with connectivity and weight information to enable subsequent routing queries to be computed very efficiently and in parallel.  To this end, we leverage Customizable Contraction Hierarchies~\cite{dibbelt2014customizable}, which further splits the preprocessing phase in two, allowing the weights of the existing links in the network to be updated with less computation time compared to doing a full topological update (where the connectivity of the links may also change).  In Algorithm~\ref{alg:parallel_aon}, the preprocessing is done first so that all threads can subsequently utilize the pre-processed network with updated weights to compute its assigned routes.

After the parallel routing and truncation step (the \textbf{for} loop in Algorithm~\ref{alg:parallel_aon}), each process's memory contains only the routes computed by the threads local to that process (implicit in $\bh_k$).  However, in order to proceed to the next algorithmic step, the impact of all routes \textit{globally} must be taken into consideration and reflected in the network flow data in every process.  A key observation is that we do not need to globally broadcast the actual routes $\bh$ computed for every vehicle trip leg, as this would be a very large amount of data to communicate.  Instead, the thread local \textit{route} flows $\bh_k$ are reduced to thread local \textit{link} flows $\bff_k$ in parallel, and then the local link flows are globally all-reduced to calculate the total link flows $\bff$ that result from all vehicle trip legs.  In this way, we avoid having to communicate any route information between parallel processes, only the resultant flows on the links themselves.

\subsection{Line search parallelization}

The next stage in Algorithm~\ref{alg:parallel_sta} after the all-or-nothing calculation is the line search to select the optimal $\alpha$ step size.  The parallelized algorithm selects the step size using $\bff$ directly instead of $\bh$ (as was done in Algorithm~\ref{alg:sta}), but is equivalent since multiplication with the incidence matrix $\bDelta$ distributes in the cost function expression, i.e.:
\begin{gather}
\bDelta \left[\bh + \alpha\cdot(\bh^{AON}-\bh)\right] = \bff + \alpha\cdot(\bff^{AON}-\bff)\textrm{, thus} \\
\alpha^\star = \argmin_\alpha \textrm{Cost\_function}\left(\bff + \alpha\cdot(\bff^{AON}-\bff)\right). \label{eq:alpha_star}
\end{gather}

If evaluated sequentially, the line search is a very computationally expensive step since it requires an iterative search where the cost function is evaluated many times for potential values of $\alpha$ until the minimum is found.  Furthermore, each single evaluation of the cost function requires summing the cost contribution of every link in the network, which is substantial for large networks with millions of links.  As a result, many computational traffic assignment implementations will simply use the method of successive averages (MSA) \cite{sbayti2007efficient} or other pre-determined step size sequences as a heuristic to select the gradient descent step size to avoid this high computational cost.  However, the MSA method has a drawback in that taking sub-optimal step sizes for each gradient descent iteration sometimes results in requiring more iterations overall for the traffic assignment to converge.  We have found that by selecting the optimal step size for each iteration, the number of gradient descent iterations may be reduced significantly for some time intervals.  Therefore, computing the line search efficiently through parallelization is an effective capability of our approach.

\begin{algorithm}
\KwData{ Network graph $G(\V, \A)$ \\
         \qquad \quad Current link flows $\bff$ \\
         \qquad \quad All-or-nothing link flows $\bff^{AON}$ \\
         \qquad \quad Current Frank-Wolfe gradient descent iteration $j$ \\
         \qquad \quad Line search iteration maximum steps $L$ \\
         \qquad \quad Convergence threshold (slope) $T_1$ \\
         \qquad \quad Convergence threshold (step size) $T_2$
    }
Initialize $\alpha = \frac{2}{2 + i}$  \tcp*{Initial guess}
\For{$l=1;\ l<L;\ l\mathrel{+}=1$ \tcp*{Quasi-Newton line search iteration}}{
  $[C(\alpha), C'(\alpha), C''(\alpha)] =$ Parallel\_cost\_function$(G, \alpha, \bff, \bff^{AON})$ \;
  \If{$C'(\alpha) < T_1$ \tcp*{Threshold 1}}{
    $\alpha^\star = \alpha$ \;
    \textbf{break}
  }
  $\alpha_{new} = \alpha - C'(\alpha) / C''(\alpha)$ \;
  Enforce $0 <= \alpha_{new} <= 1$ \;
  \If{$| \alpha_{new} - \alpha | < T_2$ \tcp*{Threshold 2}}{
    $\alpha^\star = \alpha_{new}$ \;
    \textbf{break}
  }
  $\alpha = \alpha_{new}$
}
\KwResult{$\alpha^\star$}
 \caption{Parallel\_line\_search: Quasi-Newton line search for optimal $\alpha^\star$}
 \label{alg:parallel_line_search}
\end{algorithm}

\begin{figure}[t]
  \centering
  \begin{subfigure}{0.33\textwidth}
  \centering
  \includegraphics[width=\textwidth]{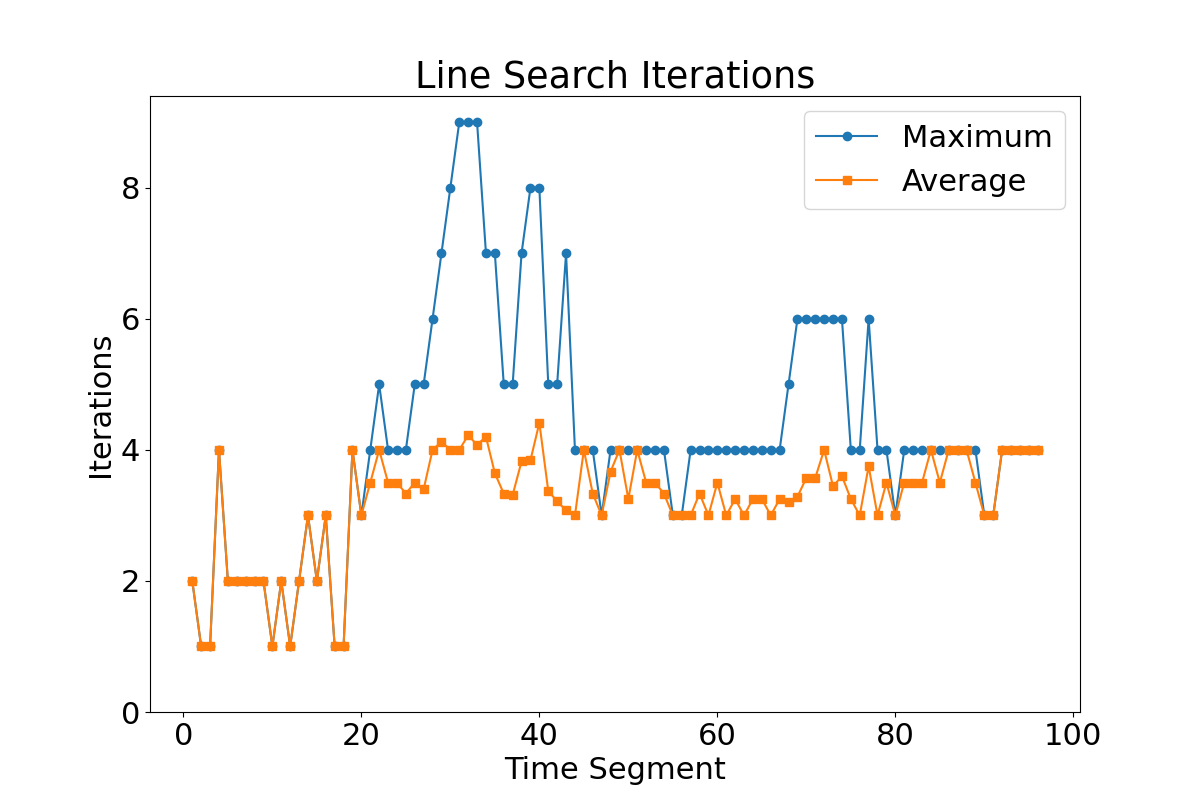} 
  \caption{}
  \label{fig:seg_ls_iters}
  \end{subfigure}
  \begin{subfigure}{0.33\textwidth}
  \centering
  \includegraphics[width=\textwidth]{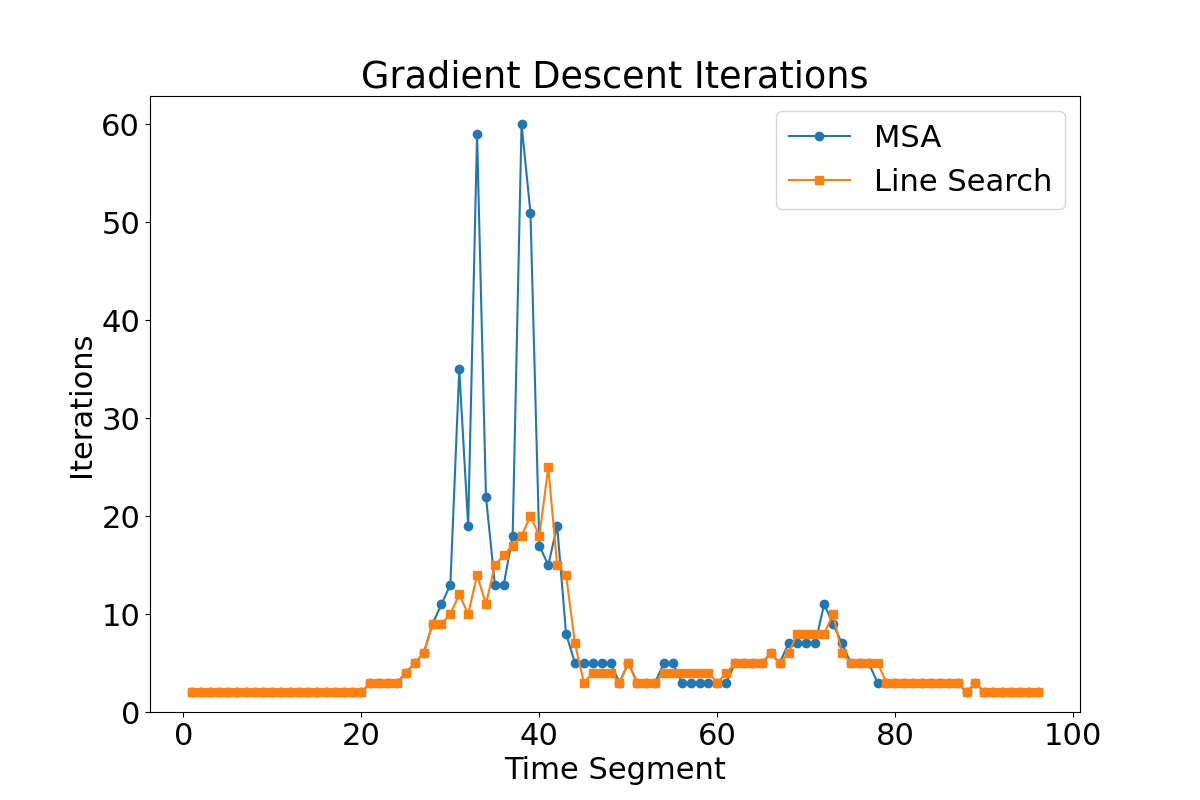} 
  \caption{}
  \label{fig:seg_iters}
  \end{subfigure}
  \begin{subfigure}{0.33\textwidth}
  \centering
  \includegraphics[width=\textwidth]{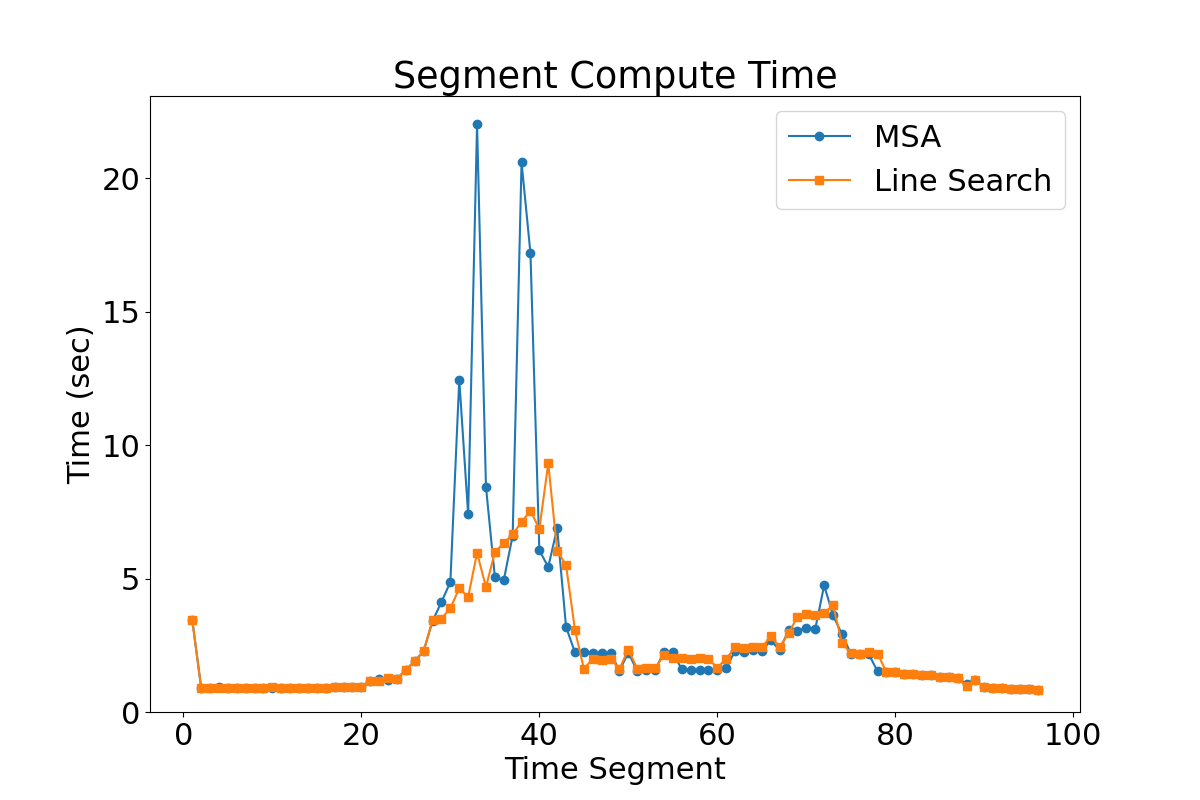} 
  \caption{}
  \label{fig:seg_times}
  \end{subfigure}
  
  \caption{a) Maximum and average number of Quasi-Newton iterations to conduct the line search over all gradient descent iterations within each time segment for the San Francisco Bay Area network model.  Comparison of b) the number of gradient descent iterations and c) the total segment compute time required using the method of successive averages (MSA, blue circles) versus using a optimal step size via Quasi-Newton method line search (orange squares).  Data is for the San Francisco Bay Area network model, and execution times are measured when running on the Cori computer with 512 cores.}
  \label{fig:msa_vs_line_search}
\end{figure}

Algorithm~\ref{alg:parallel_line_search} describes our parallel line search algorithm, which is the implementation of Equation~\ref{eq:alpha_star}.  In our algorithm, the value of $\alpha$ is constrained to the range $[0, 1]$ to ensure the solution is a convex combination of legal route assignments to prevent unrealistic (e.g. negative) flow values.  For brevity, we introduce the short-hand function $C(x)$, which is equivalent to the cost function evaluated with step size $x$ and implicit arguments $\bff$ and $\bff^{AON}$.  The algorithm uses a Quasi-Newton method to identify where $C$ is minimized (i.e. where the derivative of the $C$ equals zero).  Thus, at each step in the Quasi-Newton method, we require approximations of the first two derivatives of $C(x)$ at the current guess.  These approximations are calculated using the finite difference method by evaluating $C$ at the values: $\{x - \Delta x, x, x + \Delta x\}$, where nominally $\Delta x = 0.0001$ is chosen because it is a small fraction of the search range and works well in practice.  Then:
\begin{equation}
C'(x) \approx \frac{C(x + \Delta x) - C(x - \Delta x)}{2 \Delta x}, \quad \textrm{and} \quad
C''(x)  \approx \frac{C(x - \Delta x) - 2 C(x) + C(x + \Delta x)}{\Delta x^2}. \label{eqn:cost_derivatives}
\end{equation}
The thresholds $T_1$ and $T_2$ in the algorithm are tunable parameters to control the quality of convergence.  We have used $T_1 = T_2 = 0.0001$ in our experiments to ensure good convergence of the line search.  Figure~\ref{fig:seg_ls_iters} shows the average and maximum number of Quasi-Newton iterations required to conduct the line search for all gradient descent iterations taken within each time segment.  The average number of line search iterations ranges between 1 and 5, while the maximum is as high as 9 for the most highly congested time segments.

\begin{algorithm}
\KwData{ Network graph $G(\V, \A)$ \\
         \qquad \quad Step size $\alpha$ \\
         \qquad \quad Current link flows $\bff$ \\
         \qquad \quad All-or-nothing link flows $\bff^{AON}$ \\
         \qquad \quad Quasi-Newton step size $\Delta x$
    }
Initialize $\by_k = [0,0,0]$\;
Let $\A_k$ = this thread's partition of network links (for thread $k$)\;
Let $C(x) = $ Cost\_function$\left( \bff + x \cdot (\bff^{AON} - \bff)) \right)$ \;
\For{$a \in \A_k$}{
 $\by_k \mathrel{+}= [ C_a(x - \Delta x), C_a(x), C_a(x + \Delta x) ] $ \;
}
$\by = [C(x - \Delta x), C(x), C(x + \Delta x)] = $ Global all-reduce$(+, \by_k)$ \tcp*{Reduce values simultaneously}
$C'(x) = \frac{C(x + \Delta x) - C(x - \Delta x)}{2 \Delta x}$ \;
$C''(x) = \frac{C(x - \Delta x) - 2 C(x) + C(x + \Delta x)}{\Delta x^2}$ \;
\KwResult{$[C(x), C'(x), C''(x)]$}
 \caption{Parallel\_cost\_function: evaluate $C(x)$ and first two derivatives}
 \label{alg:parallel_cost_function}
\end{algorithm}

The key to computing the line search efficiently is evaluating $C(x - \Delta x)$, $C(x)$, and $C(x + \Delta x)$ in parallel and as a batch.  Since $C(x)$ is the sum of partial cost contributions from each link in the network (see Equation~\ref{eq:cost_function}), parallelization is achieved by partitioning the links in the network across available threads.  Algorithm~\ref{alg:parallel_cost_function} describes our approach to evaluate the cost function and its derivatives in parallel.  For each set of three cost function evaluations, each thread computes its local contributions to the three cost functions for the partition of links assigned to that thread $\A_k$.  The resulting local cost contributions are then globally all-reduced across threads to obtain the total cost function values.  In order to avoid the communication overhead of three separate global reductions for each of the three evaluations, our implementation simultaneously reduces all three values with a single vectorized all-reduce operation.  The derivatives are then estimated using the approximations in Equation~\ref{eqn:cost_derivatives}.  Due to the efficient parallel evaluation of $C(x)$ and its derivatives, we observed that even for the cases with the highest number of Quasi-Newton iterations (see Figure~\ref{fig:seg_ls_iters}), the line search completes in less than 500 milliseconds using 512 cores of the Cori supercomputer (see Section~\ref{sec:scaling} for details on Cori).

We observed the parallelized line search improves the performance of the QDTA compared to using the method of successive averages (MSA) for time segments with high levels of congestion.  Figures~\ref{fig:seg_iters} and \ref{fig:seg_times} show the impact of using a line search versus MSA on number of gradient descent iterations and total time to solution for each time segment in our experiment on the San Francisco Bay Area network (see Section~\ref{sec:scaling} for details).  The line search and MSA are roughly equivalent for most time segments with low congestion, while there is a significant improvement for the most congested time segments during the morning peak.  Furthermore, the segment compute times are highly correlated with the number of gradient descent iterations primarily because of the expensive all-or-nothing routing step required for each iteration.  By finding the optimal step size for each iteration, the number of iterations required to converge is reduced, resulting in an overall savings of 16 percent in total execution time compared to MSA.

\subsection{Network update and residual demand parallelization}

Once the line search has converged and the optimal step size $\alpha^\star$ has been identified, the link weights for the network must be updated in every process.  This step is parallelized across threads by assigning each thread a partition of the links to update the flow values and compute the new weights associated with its partition of links.  Because each process must update all of the links in the network for the subsequent routing step to perform correctly, the degree of parallelism in this step is reduced to the number of threads within each process (as opposed to across all processes).  This step corresponds to the upper blue boxes in Figure 5.

Finally, after the Frank-Wolfe algorithm has converged, the residual demand allocation must be performed to forward residual vehicles into the next QDTA time segment (Algorithm~\ref{alg:residual}).  The parallelization strategy for this algorithm partitions the traffic demand across threads in the exact manner as Algorithm~\ref{alg:parallel_aon}.  Each thread iterates over its partition ($\bd_k$) and forwards the demand that is still in transit at the end of the current time segment into the next time segment.  
Once the residual demand is computed, the non-residual demand in the next time segment is identified and combined with the residual demand in parallel, and the Frank-Wolfe optimization for the next time segment begins.  As we have described, the majority of the algorithms required for the QDTA methodology (as shown in Figure~\ref{fig:QDTAFlowDiagram}) have been parallelized to achieve high performance on distributed memory computer platforms.

\subsection{Evaluation of computational performance and parallel scalability}
\label{sec:scaling}

We use high performance computing to address the computational challenges of a traffic assignment at urban scale.  The solutions are implemented on the Cori supercomputer, a Cray X40 at the National Energy Research Scientific Computing Center (NERSC) at Lawrence Berkeley National Laboratory.  In order to evaluate the computational \textit{strong scaling} performance (improvement in program execution time for a fixed input as the number of cores is increased), we ran the QDTA algorithm for the San Francisco Bay Area network, utilizing up to 32 nodes of Cori with 1,024 cores total (2 processes per node, 16 cores per process, 2 threads per core). As we described in Section~\ref{sec:parallelization}, the computational advantage is realized through parallelization of the various algorithmic steps in \cref{fig:QDTAFlowDiagram} across many compute cores.  We segmented the 24-hour day into 96 time segments, each 15 minutes (versus the multiple hour time segments that are traditionally used).  For all runs, we solved for optimized route plans for all 19 million vehicle trips that constitute the demand $\{\bd^o(i) \ | \ \forall i\}$ over the San Francisco Bay Area network (0.5 million nodes and 1 million links).  Refer to \Cref{sec:analysis} for more details about the input network and demand.

\begin{figure}
  \centering
  \begin{subfigure}{0.49\textwidth}
  \centering
  \includegraphics[width=\textwidth]{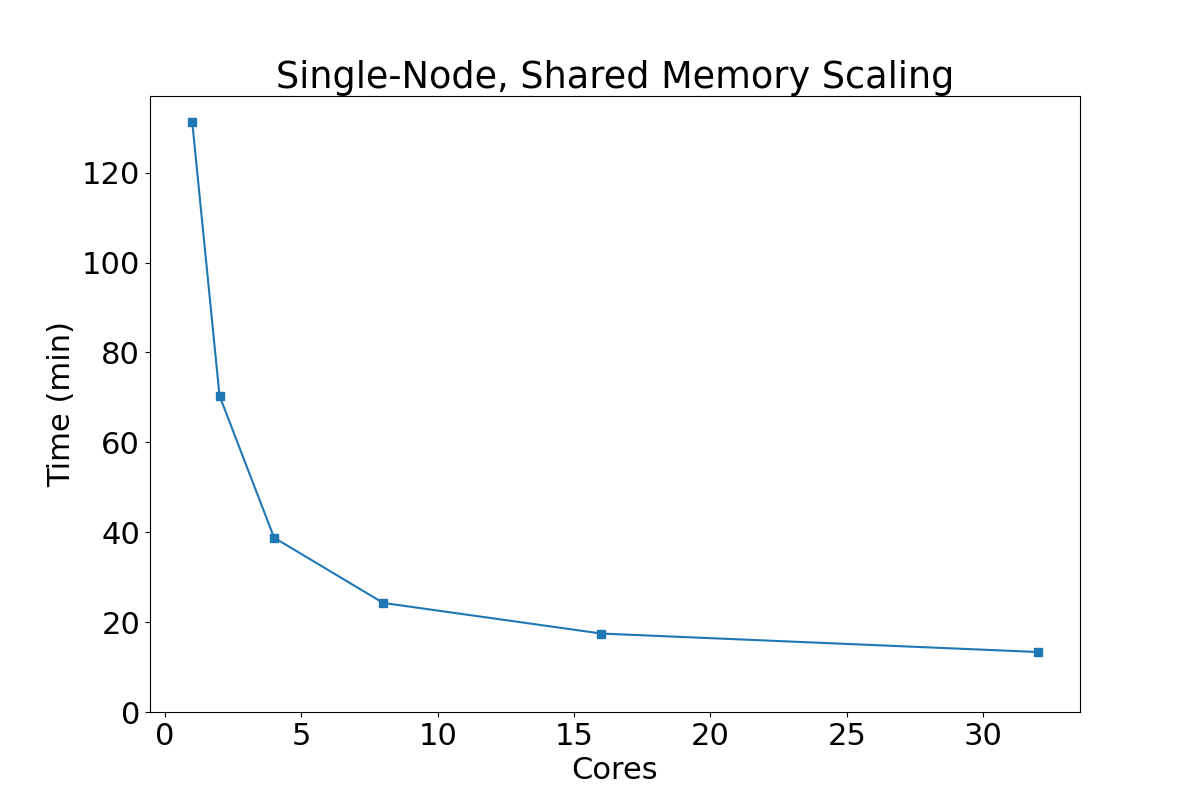} 
  \caption{}
  \label{fig:sm_scaling}
  \end{subfigure}
  \begin{subfigure}{0.49\textwidth}
  \centering
  \includegraphics[width=\textwidth]{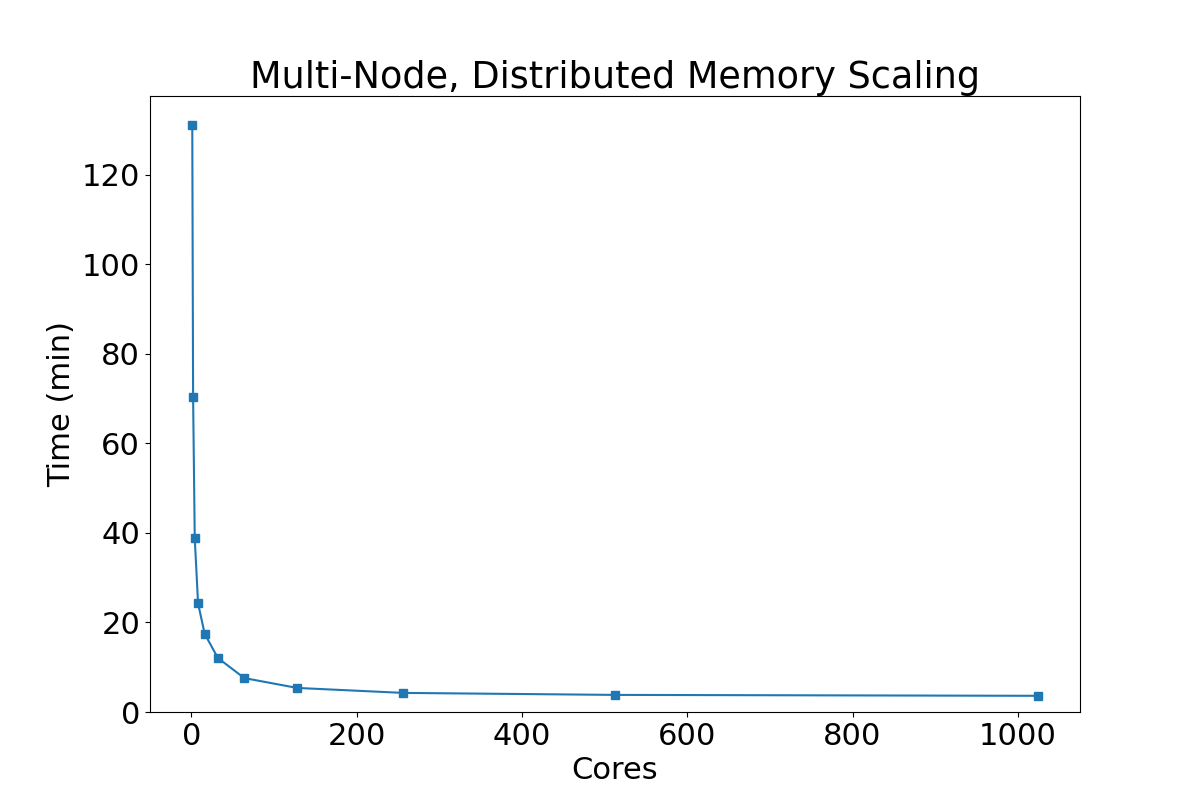} 
  \caption{}
  \label{fig:dm_scaling}
  \end{subfigure}
  \caption{Computational scaling performance for our QDTA approach as parallelism is increased.}
  \label{fig:scaling}
\end{figure}

Figure \ref{fig:scaling} shows the strong scaling performance results: a) when increase the number of cores on a single node, and b) when scaling across multiple nodes.  Running the 96 time segment QDTA on a single node, the execution time is reduced from more than two hours (131 minutes) when run on a single core to under 14 minutes utilizing all 32 cores.  This represents a parallel speedup of about 10x compared to single-core execution.  Running the QDTA on multiple nodes, we see that the execution time is reduced further to under 4 minutes when run on 32 nodes (1,024 cores), representing a total speedup of over 36x compared to the single core case.  The reason the speedup is sub-linear (less than $N$ times speed up when using $N$ times the compute cores) is due to a combination of overhead of parallelization (communication and synchronization costs between cores and processes) and the parts of the code which are not fully parallelized (resulting in Amdahl bottlenecks~\cite{Gustafson2011}).  The most notable parts that not fully parallelized are the computational routines that update the network flows and weights after each gradient descent iteration.  Because the entire network state must be replicated in each process to parallelize the expensive routing step, a fully parallel network update would require a global reduction of state the size of the entire network (millions of link weights).  To avoid this reduction, the update is partially parallelized across threads within each process.  Furthermore, the customization preprocessing step of the multi-phase routing library RoutingKit \cite{dibbelt2014customizable} has not been parallelized.

\section{Simple Network Example}
To help illustrate the QDTA model, we describe its application to a simple network with four links and five nodes and compare it with STA. For simplicity, all links are connected serially and no route choice decisions are involved in the network. The link free flow travel times and capacities are given in \cref{tab:sim_net}. We consider the demand for one hour with four time periods with 15 minutes interval. The demand is given in \cref{tab:sim_demand}. The volume delay function considered is of BPR form as shown in Equation (\ref{eq:link_travel_time}). 

\begin{table}[!htb]
\begin{minipage}{.5\linewidth}
    \centering
    \caption{Network attributes for the example}
    \label{tab:sim_net}
    \medskip
\begin{tabular}{ c c c} 
\toprule
Link & Free flow travel time(min) & Capacity(v/h)\\   
\midrule
$l_{12}$ & 10 & 200\\
$l_{23}$ & 5 & 150 \\
$l_{34}$ & 10 & 200 \\
$l_{45}$ & 10 & 200\\
\bottomrule
\end{tabular}
\end{minipage}\hfill
\begin{minipage}{.5\linewidth}
    \centering
    \caption{Demand used in the example}
    \label{tab:sim_demand}
    \medskip
\begin{tabular}{ c c c} 
    \toprule
    Time  &  OD & Demand(v/h) \\
    \midrule
    $t_{1}$ & $d_{14}$ & 175\\
    $t_{2}$ & $d_{35}$ & 50 \\
    $t_{3}$ &  & 0 \\
    $t_{4}$ &  & 0\\
    \bottomrule
\end{tabular}
\end{minipage}
\end{table}

As shown in \cref{figure:sim_qdta},for time $t_{1}$, demand $d_{14}$ traverses link $l_{12}$ and reaches node 2, but it cannot reach its destination in the same time interval and hence stored in the downstream node 3 as residual. In the next time period $t_{2}$, the new demand $d_{35}$ and the residual from previous time segment will traverse link $l_{34}$. All demand reaches its destination in time segment $t_{2}$. Time periods $t_{3}$ and $t_{4}$ have no demand and hence all trips reaches destination in the first two time interval. The total system travel time is: $ (175*10/60) + (175*6.3/60) + (225*12.4/60) + (50*10/60)  = 94$. The congested links in the network are $l_{23},l_{34} $ during time periods $t_{1}$ and $t_{2}$ respectively. 

\begin{figure}[ht!]
\centering
\begin{tikzpicture}[
roundnode/.style={circle, draw=blue!60, very thick, minimum size=10mm},
squarednode/.style={rectangle, draw=white!60, very thick, minimum size=5mm},
residualnode/.style={rectangle, draw=orange!60, very thick, minimum size=5mm},
]
\node[squarednode] (s1)at (-1,2) {\color{black} $ a) t_1$};
\node[roundnode]        (n1)                   {1};
\node[roundnode]        (n2)       [right=of n1] {2};
\node[roundnode]        (n3)       [right=of n2] {3};
\node[roundnode]        (n4)       [right=of n3] {4};
\node[roundnode]        (n5)       [right=of n4] {5};
\node[squarednode]      (d1)       [below=of n1] {175};
\node[residualnode]     (r1)       [above=of n3] {r = 175 };

\draw[->, thick ] (n1.east) to node[above]{\color{black} \footnotesize $10$ } (n2.west);
\draw[->, thick ] (n2.east) to node[above]{\color{black} \footnotesize $6.3$} (n3.west);
\draw[->, thick ] (n3.east) -- (n4.west);
\draw[->, thick ] (n4.east) -- (n5.west);
\draw[->, thick] (d1.north) -- (n1.south);
\draw[-] (r1.south) to [out=-20,in=20]node[right]{\color{white}} (n3.north);
\end{tikzpicture}
\end{figure}%

\begin{figure}[ht!]
\centering
\begin{tikzpicture}[
roundnode/.style={circle, draw=blue!60, very thick, minimum size=10mm},
squarednode/.style={rectangle, draw=white!60, very thick, minimum size=5mm},
residualnode/.style={rectangle, draw=orange!60, very thick, minimum size=5mm},
]
\node[squarednode] (s2)at (-1,2) {\color{black} $ b) t_2$};
\node[roundnode]        (N1)                   {1};
\node[roundnode]        (N2)       [right=of N1] {2};
\node[roundnode]        (N3)       [right=of N2] {3};
\node[roundnode]        (N4)       [right=of N3] {4};
\node[roundnode]        (N5)       [right=of N4] {5};
\node[squarednode]      (D1)       [above=of N3] {175};
\node[squarednode]      (D2)       [below=of N3] {50};

\draw[->, thick] (N1.east) --  (N2.west);
\draw[->, thick] (N2.east) -- (N3.west);
\draw[->, thick] (N3.east) to node[above]{\color{black} \footnotesize $12.4$ } (N4.west);
\draw[->, thick] (N4.east) to node[above]{\color{black} \footnotesize $10$ }  (N5.west);
\draw[->, thick] (D1.south) -- (N3.north);
\draw[->, thick] (D2.north) -- (N3.south);
\end{tikzpicture}
\caption{ QDTA assignment for simple network. The network is loaded in 2 time segments. $r$ represents the residual stored in the node at the end of each time period. The link travel time in minutes is noted above each link.}
\label{figure:sim_qdta}
\end{figure}
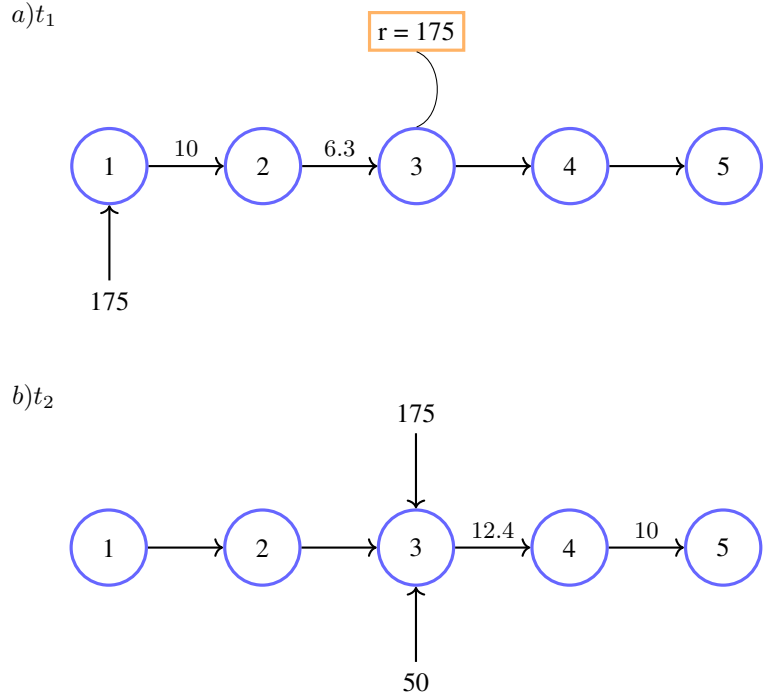

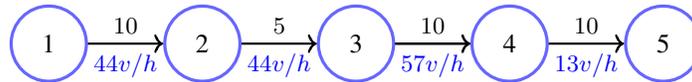
\begin{figure}[ht!]
\centering
\begin{tikzpicture}[
roundnode/.style={circle, draw=blue!60, very thick, minimum size=10mm},
squarednode/.style={rectangle, draw=white!60, very thick, minimum size=5mm},
residualnode/.style={rectangle, draw=orange!60, very thick, minimum size=5mm},
]

\node[roundnode]        (N1)                   {1};
\node[roundnode]        (N2)       [right=of N1] {2};
\node[roundnode]        (N3)       [right=of N2] {3};
\node[roundnode]        (N4)       [right=of N3] {4};
\node[roundnode]        (N5)       [right=of N4] {5};

\draw[->, thick] (N1.east) to node[above]{\color{black} \footnotesize $10$ } (N2.west);
\draw[->] (N1.east) to node[below]{\color{blue} \footnotesize $44v/h$ } (N2.west);
\draw[->, thick] (N2.east) to node[above]{\color{black} \footnotesize $5$ } (N3.west);
\draw[->] (N2.east) to node[below]{\color{blue} \footnotesize $44v/h$ } (N3.west);
\draw[->, thick] (N3.east) to node[above]{\color{black} \footnotesize $10$ } (N4.west);
\draw[->, thick] (N3.east) to node[below]{\color{blue} \footnotesize $57v/h$ } (N4.west);
\draw[->, thick] (N4.east) to node[above]{\color{black} \footnotesize $10$ }  (N5.west);
\draw[->] (N4.east) to node[below]{\color{blue} \footnotesize $13v/h$ }  (N5.west);
\end{tikzpicture}
\caption{STA assignment for simple network. Full demand is assigned at the same time and gets averaged for the entire 1 hour time duration. The link travel time in minutes and link flows in vehicles per hour is noted above and below respectively.}
\label{figure:sim_sta}
\end{figure}

The same demand is assigned using STA as shown in \cref{figure:sim_sta}. The demand is assigned as averaged over the entire 1 hour duration. The total system travel time is:
$(44*10/60)+ (44*5/60)+ (57*10/60) + (13*10/60) = 22$. None of the links are congested during any time interval in this assignment because the STA spreads the demand over the entire one-hour analysis period and is unable to capture the peak congestion that occurs on link $l_{34}$.  On the other hand, the QDTA algorithm correctly captures the peak loading of $l_{23}$ in $t_1$ as well as the residual demand resulting from $d_{14}$ that loads link $l_{34}$ at the same time that the demand $d_{35}$ enters the network in $t_2$.  We will next describe how the QDTA is able to capture similar peak loading effects over a much larger network in the following section.

\section{Application to San Francisco Bay Area Network}
\label{sec:analysis}

\subsection{Analysis of Traffic Assignment Results}

\begin{figure}[h]
\centering
\begin{subfigure}{0.90\textwidth}
\centering
\includegraphics[width=0.6\linewidth]{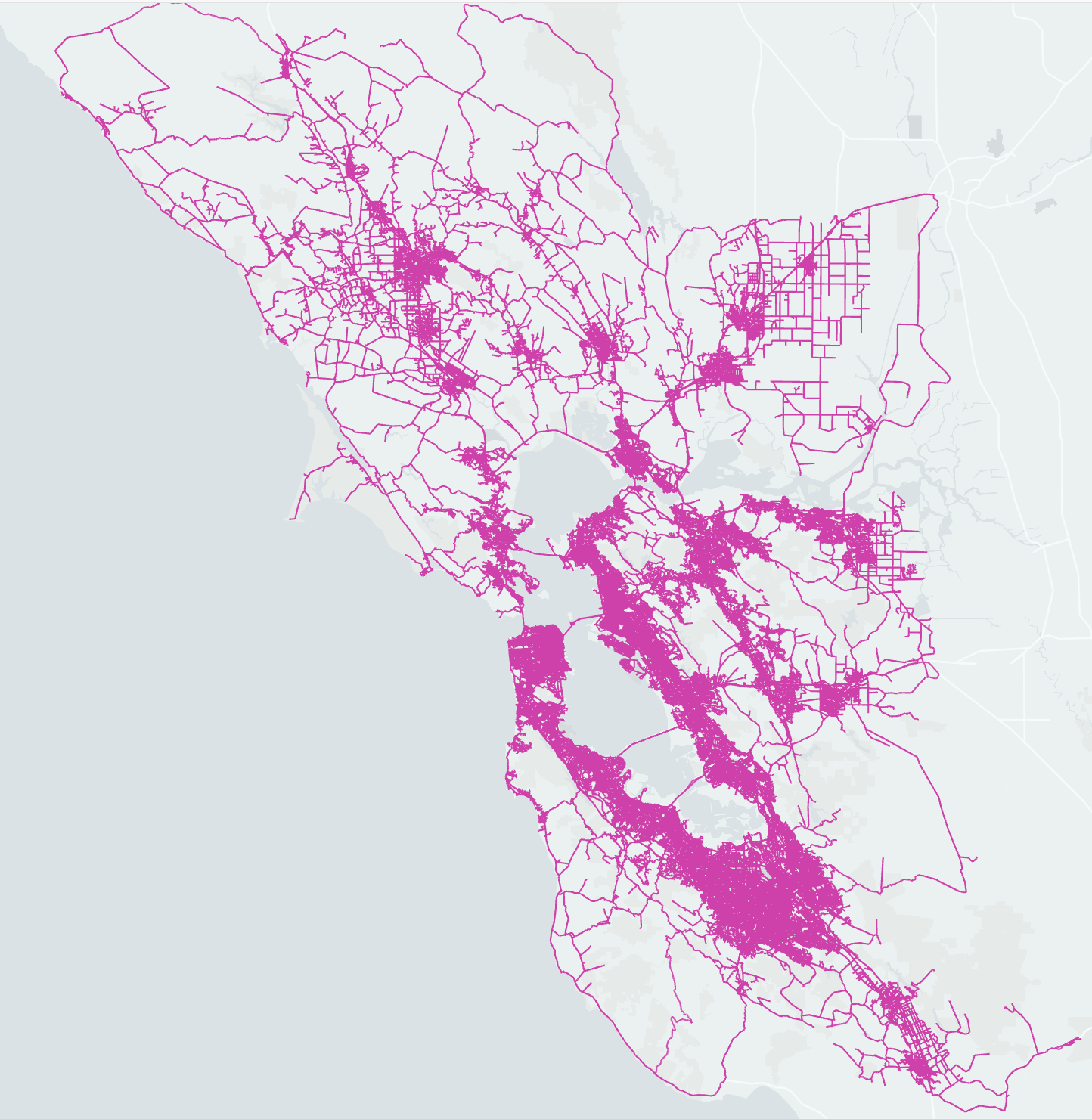} 
\caption{Bay Area Network \(\sim\)1M Links}
\label{fig:bayarea}
\end{subfigure}
\caption{Road Network for the SF Bay Area}
\label{fig:road_networks}
\end{figure}

In this section, we present an analysis of the results from applying our QDTA methodology to a large-scale urban transportation system: the San Francisco Bay Area.
For our network representation, we utilize a professional map from HERE Technologies ~\cite{here_tech}.
The HERE map divides the network links into several functional class categories, depending on their characteristics. 
Specifically, functional classes classify roads according to the speed, importance and connectivity of the road. A road can be one of five functional classes -- these are defined in \cref{table:functionalclasses}. The analysis presented here will use these functional classes to help explore the results that compare QDTA to STA. 
The road network for San Francisco is shown in \cref{fig:bayarea} and accounts for $\sim$0.5M nodes and $\sim$1M links (~\cite{here_tech}). 
We have also successfully applied our QDTA approach to the Los Angeles road network with $\sim$1M nodes and $\sim$2M links with $\sim$40M trips in a 24 hour period. For brevity, we will present results for the Bay Area only.

\begin{table}[h]
\caption{Functional Road Classes}
\begin{tabular}{|c|l|}
\hline
\rowcolor[HTML]{68CBD0} 
Functional Class & \multicolumn{1}{c|}{\cellcolor[HTML]{68CBD0}Definition}                                   \\ \hline
\rowcolor[HTML]{FFFFFF} 
1                & Allowing for high volume, maximum speed traffic movement                                  \\ \hline
\rowcolor[HTML]{FFFFFF} 
2                & Allowing for high volume, high speed traffic movement                                     \\ \hline
\rowcolor[HTML]{FFFFFF} 
3                & Providing a high volume of traffic movement                                               \\ \hline
\rowcolor[HTML]{FFFFFF} 
4                & Providing for a high volume of traffic movement at moderate speeds between neighbourhoods \\ \hline
\rowcolor[HTML]{FFFFFF} 
5                & Roads whose volume and traffic movement are below the level of any other functional class       \\ \hline
\end{tabular}
\label{table:functionalclasses}
\end{table}

We obtained demand data from the SFCTA CHAMP 6 \cite{sf-champ6.1} agent-based time-dynamical model for the Bay Area network, which uses data such as observed travel patterns, representations of the transportation system, and population and employment data to synthesize $\sim$19 million trips during a 24-hour period.
Each trip in the CHAMP model was identified with an origin and destination micro-analysis zone, which was then assigned to specific network nodes by weighting with the population density obtained from Global Human Settlement \cite{ghs}.

\begin{figure}[h]
  \centering
   \includegraphics[width=0.8\linewidth]{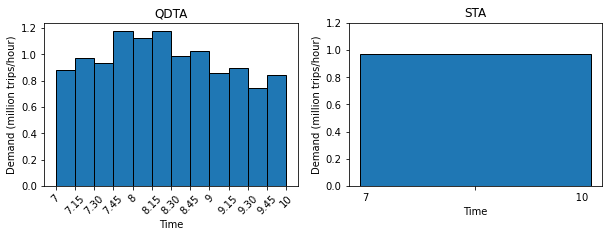}
  \caption{Temporal demand interaction and model capabilities}
  \label{fig:demandprofiling}
\end{figure}

For our analysis of QDTA, we evaluate two traffic assignment models for the San Francisco Bay Area. We compare a static traffic assignment (STA) that assigns all traffic in the 7-10 am morning peak in a single time interval, while the QDTA model divides the same period into twelve 15-minute intervals. Both cases solve for a user equilibrium traffic assignment within each time segment.  \cref{fig:demandprofiling} shows the demand profile for both cases under consideration. While STA assumes a constant demand for the entire morning peak period, QDTA enables modeling a variable demand for the same duration. 

\cref{table:systemmetricsfc} provides a comparison of the system level metrics using vehicle miles travelled (VMT), average volume over capacity (VOC) ratios, and vehicle hours of delay (VHD) categorised by functional class. Only links with positive flow are included in the analysis.  Additionally, \cref{table:congestedmetrics} provides a comparison between STA and QDTA for the \textbf{congested} portion of the network, where a link is considered congested if its volume over capacity ratio is greater than or equal to 1.0.  For QDTA analysis, congestion is calculated for 7.45-8 AM time period when the demand is the highest. 

\begin{table}[h]
\centering
\caption{System Level Metrics (all roads)}
\label{table:systemmetricsfc} 
  \begin{tabular}{|P{2cm}|P{2cm}|P{2cm}|P{2cm}|P{2cm}|P{2cm}|P{2cm}|} 
    \hline
    \multirow{2}{*}{\textbf{Category}}& \multicolumn{2}{|c|}{\textbf{VMT (in millions)}} & \multicolumn{2}{|c|}{\textbf{Average VOC}} & 
    \multicolumn{2}{|c|}{\textbf{VHD (in thousands)}}\\
    \cline{2-3} \cline{4-5} \cline{6-7}
  & STA  & QDTA &  STA & QDTA &  STA & QDTA \\ 
    \hline
    FC2 & 16.44 & 18.11 	& 0.52	& 0.69 & 17.57 & 96.84 \\ 
    
    FC3 & 4.67 & 5.01 &	0.23 &	0.30 & 2.67 & 14.43  \\ 
    
    FC4 & 4.96	& 5.25 & 0.14 &	0.17 & 1.13 & 6.42 \\ 
    
    FC5 & 2.43 & 2.60 &	0.03 &	0.48 & 0.38 & 17.78  \\ 
    \hline
    Total & 28.51 & 30.98 & - & -  & 21.95 & 135.49 \\
    \hline
  \end{tabular}
\end{table}

\begin{table}[h]
\centering
\caption{Congested Network Metrics (roads with VOC >= 1.0)}
\label{table:congestedmetrics} 
  \begin{tabular}{|P{2cm}|P{2cm}|P{2cm}|P{2cm}|P{2cm}|P{2cm}|P{2cm}|} 
    \hline
    \multirow{2}{*}{\textbf{Category}}& \multicolumn{2}{|c|}{\textbf{Length (km)}} & \multicolumn{2}{|c|}{\textbf{VMT (in millions)}} & \multicolumn{2}{|c|}{\textbf{Average VOC}}\\
    \cline{2-3} \cline{4-5} \cline{6-7}
  & STA  & QDTA &  STA & QDTA & STA & QDTA \\ 
    \hline
    FC2 & 131 &	553 &	1.77&	7.56&	1.15&	1.26 \\ 
    
    FC3 & 38 &	88 &	0.17 &	0.42	&1.18&	1.28 \\ 
    
    FC4 & 14 &	56 &	0.04&	0.15 &	1.19&	1.21 \\ 
    
    FC5 & 5 &	28 &	0.01&	0.06 &	1.17&	1.25 \\ 
    \hline
    Total & 188  & 725 & 2.0 & 8.19 & - & -  \\
    \hline
  \end{tabular}
\end{table}

\begin{figure}[h]
  \centering
  \includegraphics[width=.5\linewidth]
    {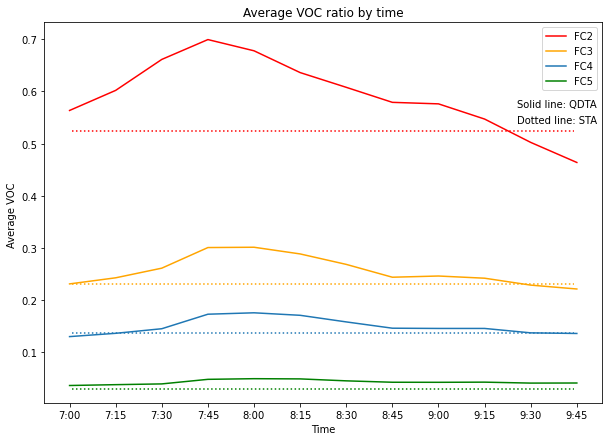}\hfill
  \includegraphics[width=.5\linewidth]
    {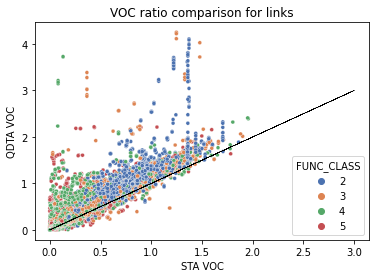}
  \caption{Congestion profiling for QDTA and STA. Figure on the left shows the average VOC ratio for all links over time categorised by functional classes. The figure on right is a scatter plot showing the VOC ratio comparison for every link in the network for the 2 models at 7:45-8 am.}
  \label{fig:metric_comparison}
\end{figure}

The key distinction between the two models is how they estimate the variation and magnitude of network congestion patterns. As seen in \cref{table:congestedmetrics}, QDTA predicts more congested roadway (length), and higher VMT and average VOC on the congested roadways compared to STA.  These data are in accordance with our expectation that non-uniform demand distributions result in higher travel times due to the convexity properties associated with link performance functions.
Furthermore, STA (by definition) is not able to recreate any temporal congestion dynamics, whatsoever.  \cref{fig:metric_comparison} shows the VOC over time by functional class (left) and VOC comparison for the two models by functional class (right). For all classes of roads the QDTA predicts congestion dynamically over time whereas STA either overestimates or underestimates the values irrespective of the demand dynamics. This difference is highly pronounced for FC2 where STA significantly underestimates the peak hour congestion. As FC2's are high capacity highways and freeways, static average demand is not able to capture the evolving congestion dynamics which is captured by QDTA. 

\begin{figure}[h]
  \includegraphics[width=.5\linewidth]
    {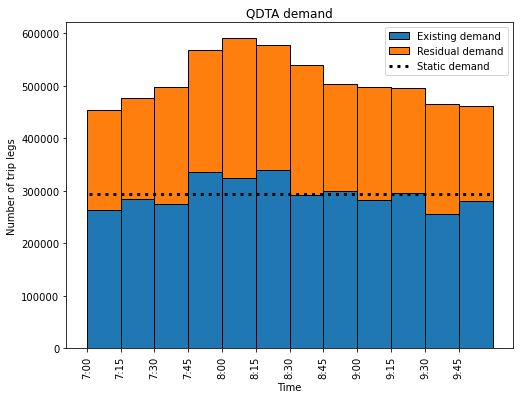}\hfill
  \includegraphics[width=.5\linewidth]
    {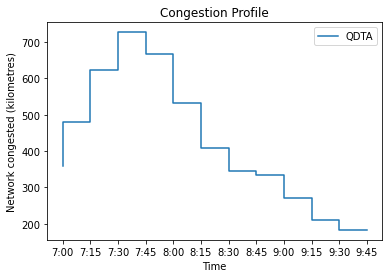}
  \caption{QDTA demand modeling with residuals (left). The length of network congested over time. A link is considered congested if the VOC ratio is greater than or equal 1.}
  \label{fig:residual_d}
\end{figure}

\begin{figure}[h]
  \includegraphics[width=.43\linewidth]
    {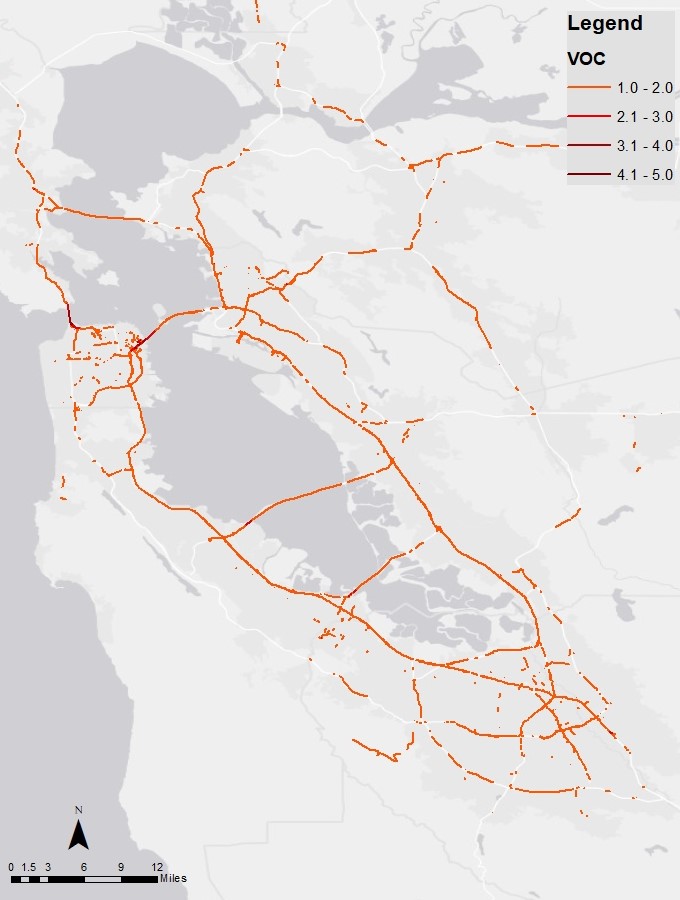} \hfill
  \includegraphics[width=.46\linewidth]
    {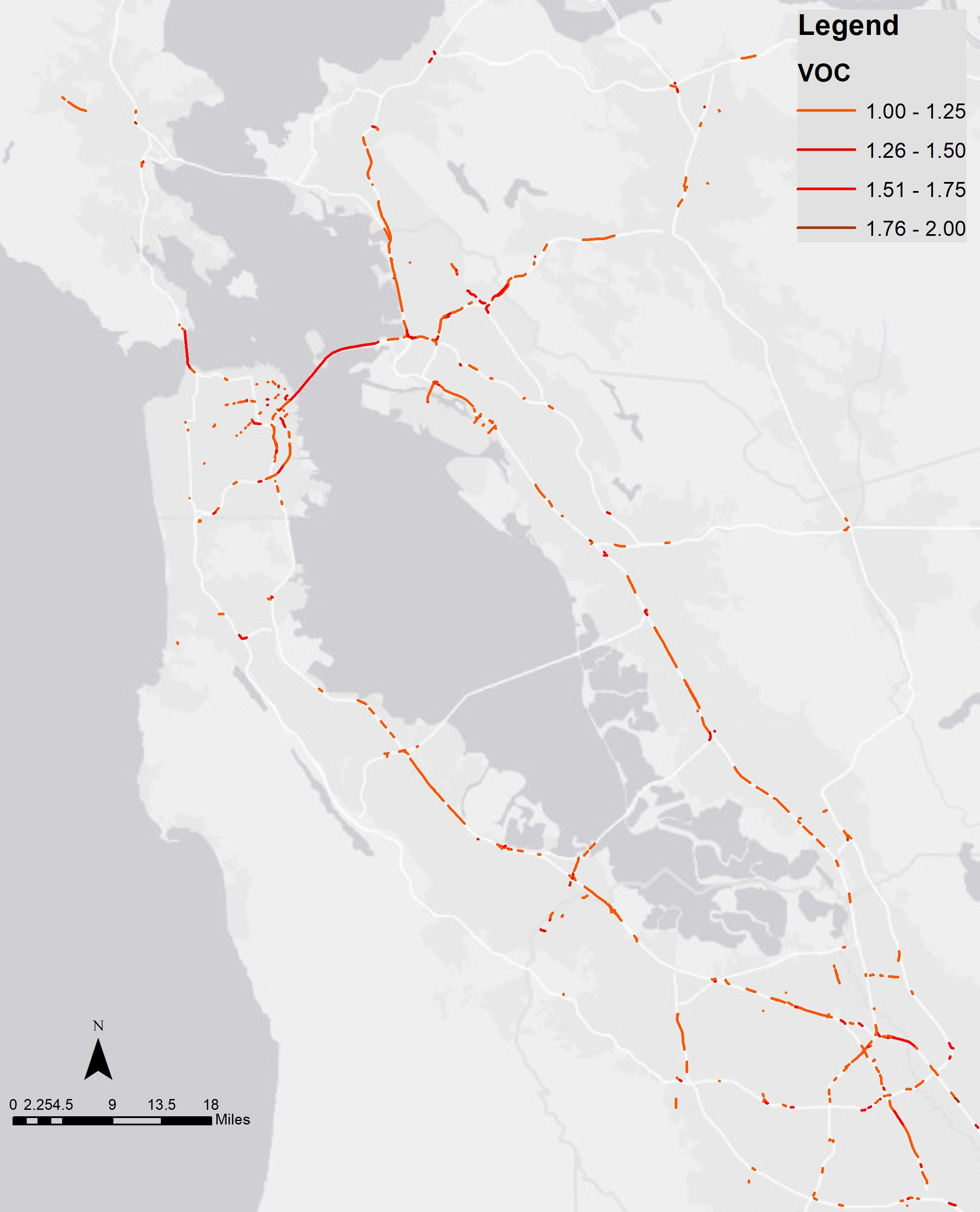}
  \caption{VOC ratios for congested links for QDTA (left) and STA (right) is shown for time period 7.45 - 8 AM. The links that are predicted to be congested in STA has even higher congestion in QDTA. STA produces 188 km of congested network in comparison to QDTA with 720 km of congested network. Only links with VOC greater than or equal to 1 is shown.}
  \label{fig:voc}
\end{figure}

It can also be seen from \cref{table:systemmetricsfc} that the total system delay is significantly higher in QDTA versus STA. This is due to the dynamic demand distribution and the modeling capability in QDTA that allows interaction of demand between time intervals, thus allowing for a more realistic modeling of traffic. 
\cref{fig:residual_d} shows the dynamic variation in the number of trip legs over time in 15 minute intervals from 7am to 10am.  For STA, the system is modeled as if the demand (3.5M trips whose start times are within the 7-10am range) are spread evenly across all three hours, resulting in an average of about 300k trips per 15 minute interval.  As described in \cref{sec:algorithms}, for QDTA the total number of trips modeled in an interval is composed of two parts: the original (existing) demand $\bd^o(i)$ and the residual demand $\bd^r(i)$ carried over from previous intervals.  While the sum of the existing and residual trips appears to be much larger than the static demand case, in reality each trip only contributes a fraction of its entire route to each time segment due to the route truncation mechanism.
For trips that span multiple time intervals, different (non-overlapping) segments of the trip's path are loaded onto network in each time interval.
Thus, in QDTA each trip leg only contributes flow to the links it traverses within the time segment, whereas for STA every leg contributes flow to the entire route. For the time interval 7.45-8 AM, STA has 188 km of congested network in comparison to QDTA with 725 km of congested network. \cref{fig:voc} shows the Bay area network with congestion locations for the two modeling cases.

\subsection{Validation}
\begin{figure}[t!]
  \centering
  \includegraphics[width=.43\columnwidth]
    {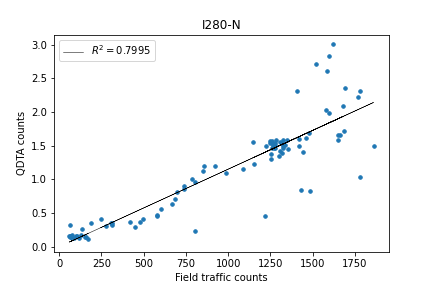}
  \includegraphics[width=.43\columnwidth]
    {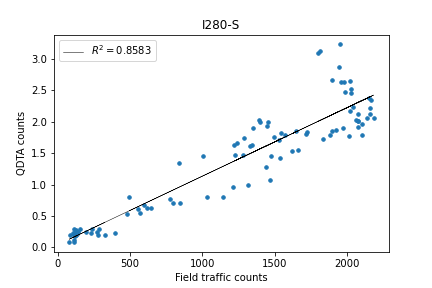}
  \includegraphics[width=.43\columnwidth]
    {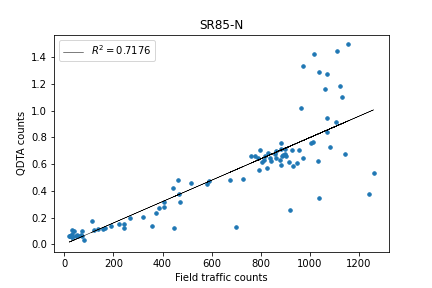}
  \includegraphics[width=.43\columnwidth]
    {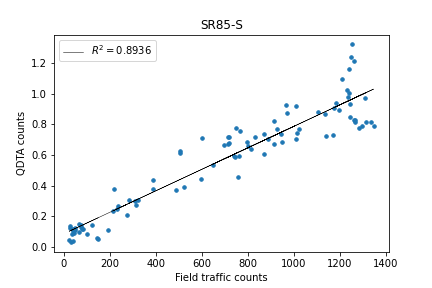}
  \includegraphics[width=.43\columnwidth]
    {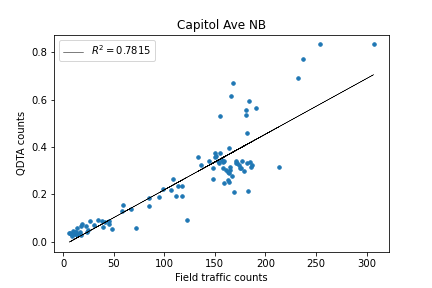}
  \includegraphics[width=.45\columnwidth]
    {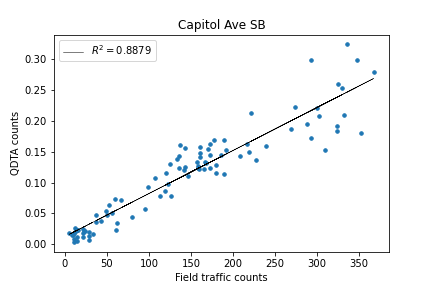}
  \includegraphics[width=.43\columnwidth]
    {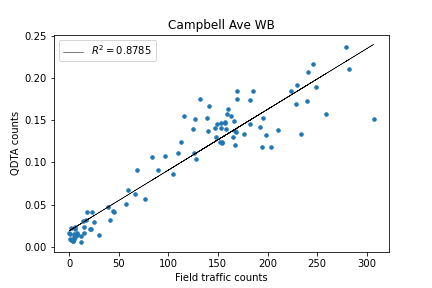}
  \includegraphics[width=.43\columnwidth]
    {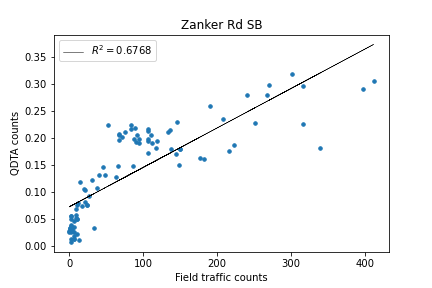}
  \caption{Validation of traffic counts in 15 minute increments for links in different functional classes. All links except one have satisfactory R$^{2}$ of greater than 0.7}
  \label{fig:stage1a}
\end{figure}

Validation was performed for QDTA results to test the effectiveness of representing the real world traffic environment. We conducted validation for traffic volume, speed and system metrics using multiple data sources. Stage 1 of the validation procedure involves checking the traffic counts for eight link corridors. The traffic count for each link was compared against the field data for the entire day in 15 minutes increments. The field data for city roads and highways were collected from the city of San Jose and Caltrans PeMS website \cite{pems} respectively for the year 2019. Each corridor provided information regarding traffic volumes and speeds by time of the day and direction. A coefficient of determination (R$^2$) of 0.7 is typically used as a satisfactory criterion for link count checks. The \cref{fig:stage1a} shows R$^{2}$ values for the eight corridors under consideration. The modeled corridors indicate a close match with the field data with the lowest R$^2$ value observed being 0.67 for Zanker Road. 

\begin{figure}[t!]
  \centering
  \includegraphics[width=.4\columnwidth]
    {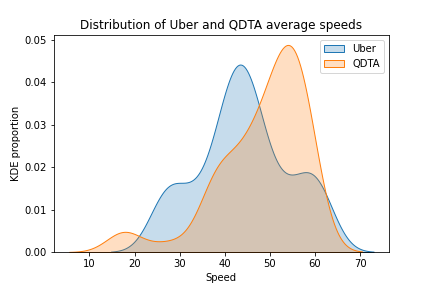} \qquad
  \includegraphics[width=.4\columnwidth]
    {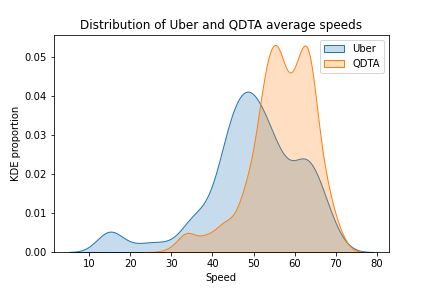}
  \caption{Kernel density plot comparing  QDTA model and Uber speed distributions at 60 mph (left) and 70 mph (right).}
  \label{fig:stage2a}
\end{figure}

\begin{figure}[t!]
  \centering
  \includegraphics[width=.4\columnwidth]
    {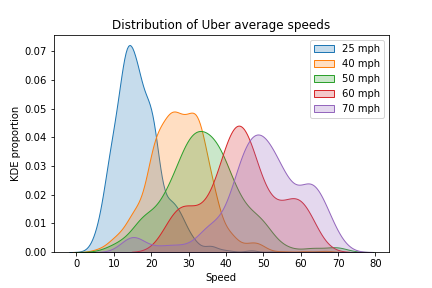} \qquad
  \includegraphics[width=.4\columnwidth]
    {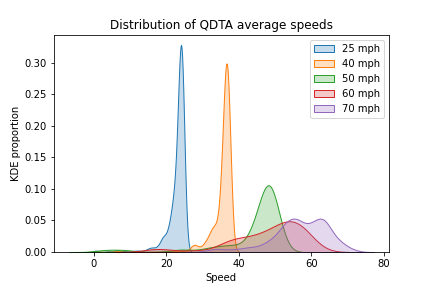}
  \caption{Distribution of average Uber (left) and QDTA model (right) speeds across specific speed limits between 8-9am in the morning shown using a kernel densities.}
  \label{fig:stage2b}
\end{figure}

Stage 2 is speed comparison with Uber Movement speed data for San Francisco region for Q4, 2019 \cite{noauthor_uber_nodate}. Links from Uber network were matched to our network for 139,495 links (20\% of total). The speeds were compared for 8 am to 9 am for different speed limits. Figure~\ref{fig:stage2a} shows the speed distributions from QDTA results and Uber on links with 60 mph and 70 mph speed limit.  Figure~\ref{fig:stage2b} shows the average speeds from Mobiliti and Uber across all speed limits. We believe the observed discrepancies in the speed distributions may be improved in future work with the addition of real world traffic signal location and timing data and further refinement of the link flow congestion and timing models. 

Final stage of validation includes system level metrics comparisons, network validation, and error checking. Model visualization is used to check for unusual activities in traffic flows and odd roadway network attributes. Error checking and model verification consist of several smaller tasks such as checks for link geometry and connectivity, number of lanes, speeds, ramps and intersection geometry.  Since our travel demand data was obtained from SFCTA, which conducts their own validation, we did not conduct additional behavior checks.  We conducted system metric checks for VMT and total demand and validated them against the 2017 Environmental Impact report for the Bay Area \cite{noauthor_environmental_nodate} in Table~\ref{tab:stage3}.

\begin{table}[h]
\centering
\caption{System level metrics validation}
\label{tab:stage3} 
  \begin{tabular}{|P{3cm}|P{3cm}|P{3cm}|P{3cm}|} 
    \hline
    \textbf{Metric} & \textbf{QDTA } & \textbf{Field Data} & \textbf{Relative Error(\%)}\\
    \hline
    VMT & 150,634,403 & 158,406,800 & -4.9 \\
Daily Trips & 19,167,301 & 21,227,800 & -10 \\
    \hline
  \end{tabular}
\end{table}

\section{Conclusions}
In this paper, we have presented a quasi-dynamic traffic assignment methodology to capture temporal dynamics in large-scale transportation networks and described its parallelization to run efficiently on distributed-memory high performance computing systems.  Two key mechanisms, route truncation and residual demand, were implemented to provide more realistic demand profiles as the dynamic assignment interval is reduced and a greater percentage of vehicle trips span across multiple time segments.  Our approach divides the simulated day into 15 minute intervals and utilizes a modified static traffic assignment within each interval to assign the active traffic present in the network.  The QDTA assignment step differs from a traditional STA in that the assigned routes are truncated to fit in the active time interval in each Frank-Wolfe iteration so that the resulting solution only includes traffic that occurs within the current interval.  Furthermore, residual demand for each time interval is calculated based on estimated travel time on the links and then carried over to the next time interval.  The combination of these techniques resolves traffic that is resident in the network for a longer time horizon than a single static assignment period and captures dynamic network behavior across multiple time segments.  The model can be interpreted as an time expanded network approach where the expansion of the network represents the evolution with respect to time.

We have also described how the quasi-dynamic traffic assignment algorithm can be parallelized for efficient execution on high-performance distributed-memory computing platforms.  The algorithm is parallelized through a combination of partitioning trip legs and network links across available compute threads to speed up the calculation of shortest paths, optimization cost functions, residual trip legs, and other program data.  We described our parallelized line search algorithm which enables the identification of the optimal step size for each gradient descent iteration using Newton's method, taking less than 500 milliseconds for a network of 1 million links.  Using the optimal step size provides a reduction of 16 percent in total execution time compared to using the method of successive averages, due to a decrease in gradient descent iterations required for convergence.  We demonstrated that a quasi-dynamic traffic assignment of the San Francisco Bay Area (19 million trip legs, 0.5 million nodes, and 1 million links) using 96 15-minute time segments runs in under 4 minutes on 1,024 cores of the Cori supercomputer at NERSC, corresponding to a speedup of 36x compared to single core performance.

Finally, we presented an analysis of the traffic assignment results across functional classes, illustrating how the QDTA more accurately resolves the increased congestion patterns and dynamic behavior of the traffic system compared to a static traffic assignment approach, especially under peak congestion conditions.  We presented a validation of the QDTA assignment counts and speeds compared to field data from CalTrans, San Jose, and Uber Movement, showing field count correlation values of 0.68 or greater.  Future work includes evaluating a variety of optimization objective functions, include fuel optimization and system level optimizations for both travel time and fuel use.  The QDTA results can be integrated into our computational platform Mobiliti~\cite{chan2018mobiliti}, which also provides parallel discrete event traffic simulations for urban-scale regions to conduct hypothetical demand, infrastructure, and advanced traffic control scenario evaluations.
We hope to also develop surrogate models using the results from the HPC implementation such that this capability can be made more widely available. 

\section*{Acknowledgements}
This report and the work described were sponsored by the U.S. Department of Energy (DOE) Vehicle Technologies Office (VTO) under the Big Data Solutions for Mobility Program, an initiative of the Energy Efficient Mobility Systems (EEMS) Program. The following DOE Office of Energy Efficiency and Renewable Energy (EERE) managers played important roles in establishing the project concept, advancing implementation, and providing ongoing guidance: David Anderson and Prasad Gupte. 
This research used resources of the National Energy Research
Scientific Computing Center, a DOE Office of Science User Facility
supported by the Office of Science of the U.S. Department of Energy
under Contract No. DE-AC02-05CH11231. 
\bibliographystyle{myunsrt}  
\bibliography{references.bib}

\end{document}